\documentclass[times]{aastex701}
\usepackage{amsmath}
\usepackage[utf8]{inputenc} 
\usepackage{CJK}

\begin{document}

\title{What Does 2D Coronal Seismology Measure?}

\begin{CJK*}{UTF8}{gbsn}
\author[orcid=0000-0002-4973-0018]{Zihao Yang (杨子浩)}
\affiliation{High Altitude Observatory, National Center for Atmospheric Research, Boulder, CO, USA, 80301}
\email[show]{yangzihao96@gmail.com}  

\author[orcid=0000-0003-1027-0795]{Yuhong Fan}
\affiliation{High Altitude Observatory, National Center for Atmospheric Research, Boulder, CO, USA, 80301}
\email{yfan@ucar.edu}  

\author[orcid=0000-0002-8439-6166]{Giuliana de Toma}
\affiliation{High Altitude Observatory, National Center for Atmospheric Research, Boulder, CO, USA, 80301}
\email{detoma@ucar.edu}

\author[orcid=0000-0001-9831-2640]{Sarah Gibson}
\affiliation{High Altitude Observatory, National Center for Atmospheric Research, Boulder, CO, USA, 80301}
\email{sgibson@ucar.edu}

\author[orcid=0000-0001-5850-3119]{Matthias Rempel} 
\affiliation{High Altitude Observatory, National Center for Atmospheric Research, Boulder, CO, USA, 80301}
\email{rempel@ucar.edu}

\begin{abstract}
Two-dimensional (2D) coronal seismology based on coronagraphic imaging spectroscopy has emerged as a promising method for mapping the coronal magnetic field, but the physical meaning of its inferred quantities under optically thin line-of-sight (LOS) integration remains uncertain. We investigate this problem by synthesizing coronagraphic observables from a global wave-driven magnetohydrodynamic model with a long LOS extent and a structured corona, and by applying the same analysis procedures used for observations. We compare the diagnosed density, wave propagation direction, phase speed, and magnetic-field strength with different LOS-weighted reference quantities from the model. We find that the measured density and the wave propagation direction are both well represented by LOS emissivity-weighted quantities. In contrast, the phase speed is not controlled by emissivity weighting alone: the LOS distribution of wave velocity amplitude also plays an important role. As a result, the inferred magnetic field is better represented by combining the emissivity-weighted density with an amplitude-dependent effective speed. Emissivity weighting nevertheless remains a reasonable approximation, and a useful baseline, over much of the field of view, especially where the LOS does not cross strongly structured plasma-sheet-like regions. These results clarify the physical interpretation of 2D coronal seismology and provide guidance for future coronal magnetic-field tomography and model-constrained applications.
\end{abstract}


\section{Introduction}

The coronal magnetic field plays a fundamental role in shaping the structure and dynamics of the solar atmosphere, yet routine measurements over a large field of view (FOV) remain difficult. In recent years, important progress has been made in coronal magnetic-field measurements. For example, using the Daniel K. Inouye Solar Telescope, \citet{2024SciA...Schad} obtained a coronal magnetic-field map based on Zeeman-sensitive polarized coronal emission, and \citet{2026ApJ..Wraback} inferred the magnetic-field strength in coronal cavities using the same method. These studies demonstrate the growing capability of spectropolarimetry for obtaining coronal magnetograms, although routine measurements over extended FOVs remain challenging, motivating the proposed Coronal Solar Magnetism Observatory \citep[COSMO,][]{2016JGRA..121.7470T}, which would obtain global synoptic observations of Zeeman-sensitive coronal emission lines.

A complementary approach is provided by coronal seismology. In its traditional form, coronal seismology infers plasma parameters, including magnetic-field strength, by combining observed wave or oscillation properties with magnetohydrodynamic wave theory \citep[e.g.,][]{2001A&A...372L..53N,2022ApJ...930...Gao}. Such measurements have been successfully applied to individual coronal structures, such as loops and plumes, and have provided valuable estimates of the magnetic field in localized regions. However, because traditional coronal seismology usually relies on isolated wave events or oscillating structures, it does not readily provide routine large-scale maps of the coronal magnetic field. A two-dimensional (2D) coronal seismology technique has been developed based on observations from the Coronal Multi-channel Polarimeter \citep[CoMP;][]{2008SoPh..247..411T} and the Upgraded Coronal Multi-channel Polarimeter \citep[UCoMP;][]{2016JGRA..UCoMP}. CoMP and UCoMP are coronagraphs that observe the coronal Fe \textsc{xiii} infrared emission lines with tunable-filter spectral sampling over an extended off-limb FOV. Their Doppler-velocity observations reveal ubiquitous propagating transverse-wave signatures in the corona \citep[e.g.,][]{2007Sci...317.1192T,2009ApJ...Tomczyk,2015NatCo...Morton}. It has been proposed that the phase speed of these transverse waves is related to the coronal density and magnetic-field strength, and can therefore be used to measure the coronal magnetic field over an extended FOV \citep[e.g.,][]{2017A&A...Long,2020Sci...369..694Y}. In this technique, the phase speed and wave propagation direction are measured through wave tracking of the Doppler-velocity signals, while the density is diagnosed from the Fe \textsc{xiii} line ratio. Together, these quantities provide proxies for inferring the magnetic-field strength and direction. The first global map of the off-limb coronal magnetic field was obtained using this technique \citep{2020Sci...369..694Y,2020ScChE..63.2357Y}. More recently, using continuous UCoMP observations, \citet{2024Sci...386...76Y} obtained more than 100 groups of coronal magnetograms, representing a first step toward routine coronal magnetic-field measurements.

As with most optically thin coronal observations, the quantities inferred by 2D coronal seismology are affected by line-of-sight (LOS) integration. This applies not only to the density diagnosed from the Fe \textsc{xiii} line ratio, but also to the phase speed and wave propagation direction derived from Doppler-velocity wave tracking, and therefore to the inferred magnetic field. In a structured corona, LOS-integrated observables may not correspond directly to local plasma parameters, but rather to effective quantities resulting from LOS weighting over multiple emitting structures.

Previous studies have discussed LOS-integration effects of coronal observations. For example, \cite{2009A&A...495..Young} noted that densities diagnosed from optically thin line ratios represent emission-weighted averages, while \cite{2023SoPh..298..Liu} and \cite{2013PASJ...65S..14I} suggested that magnetic-field strengths inferred from spectropolarimetric and radio observations should likewise be interpreted as LOS emissivity-weighted averages. \cite{2026NatAs..10...42M} further suggested that Doppler velocities observed in optically thin coronal lines are also emissivity-weighted along the LOS. However, the physical meaning of the quantities inferred by 2D coronal seismology remains unclear.

These considerations raise two closely related questions: how reliable is 2D coronal seismology in practice, and what do the physical quantities derived from this technique actually represent? Recent studies have begun to examine both the validation of the method and the interpretation of its LOS-integrated observables. For example, \cite{2024Sci...386...76Y} compared coronal magnetic-field maps measured from UCoMP observations with reference quantities calculated from global MHD models constructed from observational boundary conditions. They found that the measurements are, in general, consistent with LOS emissivity-weighted averages in the models, especially in low-to-mid latitude regions. However, because such models cannot be expected to reproduce the observed corona exactly, comparisons of this kind are not fully self-consistent for establishing the precise physical meaning of the measured quantities, and should therefore be regarded as approximate guidance.

A more self-consistent approach is provided by forward modeling, in which synthetic observables are generated from a model and analyzed in the same way as real observations, allowing the inferred quantities to be compared directly with reference quantities calculated from the model itself. Using an MHD model with a fixed magnetic-field configuration, \cite{2018ApJ...856..Magyar} synthesized emission and Doppler-velocity signals and applied coronal seismology to infer the magnetic-field strength. Following a similar approach, but including gravitational stratification, \cite{2025RAA....Gao} also compared the magnetic-field strength inferred from 2D coronal seismology with the ground truth in the model. These studies showed that, under their assumed model setups, the technique can in general recover an accurate magnetic-field strength, although the models were still highly simplified and idealized.

A recent forward-modeling study further validated the technique using a realistic coronal MHD simulation (MURaM), in which the wave signatures are naturally and self-consistently produced by convective motions in the model domain, and showed that, under relatively simple LOS conditions, the inferred magnetic field can be reasonably approximated by a LOS emissivity-weighted quantity \citep{2026ApJS..283....3Y}. This represents a more realistic coronal environment, but the FOV of the simulation is still limited, the LOS integration length remains much shorter than in real observations, and the viewing angle was chosen such that most structures were well aligned with the plane-of-sky (POS), thereby reducing the effective complexity of LOS structuring. It therefore remains unclear how the quantities measured by 2D coronal seismology should be interpreted under more realistic conditions, particularly when the LOS is longer and crosses more complex coronal structures, as is often the case in real observations.

In this work, we investigate what physical quantities are actually probed by 2D coronal seismology using a global wave-driven MHD model with a much longer LOS extent and a more structured corona than in previous forward-modeling studies. We synthesize observables following the exact sampling approach of CoMP and UCoMP, apply the same analysis procedures used in observations, and compare the measured density, wave propagation direction, phase speed, and magnetic field with different model reference quantities. In this way, we assess which diagnostics remain well described by LOS emissivity-weighted averages, where additional weighting effects become important, and under what coronal conditions the commonly adopted emissivity-weighting interpretation remains valid. The resulting physically motivated LOS-weighted relationships are not only important for interpreting current 2D coronal seismology measurements, but may also be useful for future 3D coronal magnetic-field tomographic inversions and for constraining and improving existing coronal magnetic-field models.

\section{Forward Modeling Framework} 
\subsection{2D Coronal Seismology}\label{sec:cs}
CoMP and UCoMP are imaging-spectroscopic coronagraphs that observe the coronal Fe \textsc{xiii} infrared lines at 1074.7 and 1079.8 nm with a tunable-filter sampling a few discrete wavelength positions. From these measurements, spectral parameters such as peak intensity, Doppler velocity, and line width can be derived. The Doppler velocity observations from both instruments have revealed ubiquitous propagating wave signatures that have been extensively studied over the past two decades.

In 2D coronal seismology, these propagating wave signatures are used to measure the coronal magnetic-field strength and direction. The observed transverse disturbances are interpreted as highly incompressible kink-like transverse wave motions \citep[e.g.,][]{2015NatCo...Morton,2019NatAs...Morton}. Under coronal conditions and at the instrumental resolution of CoMP/UCoMP, the measured phase speed ($v_\text{ph}$) is closely related to the coronal density ($\rho$) and magnetic-field strength ($B$) \citep[e.g.,][]{2017A&A...Long,2020Sci...369..694Y,2024Sci...386...76Y} through
\begin{equation}\label{eq:kinkeq}
    B = v_\text{ph}\sqrt{\mu_0 \left< \rho \right>},
\end{equation}
where $\mu_0$ is the magnetic permeability and $\left< \rho \right>$ denotes the average density associated with unresolved plasma structures within one observed pixel.

In practice, the analysis involves three steps: density diagnostics from the Fe \textsc{xiii} line ratio, determination of the wave propagation direction from Doppler-velocity wave tracking, and measurement of the wave phase speed along the tracked wave path. Together, these quantities provide access to both the direction and strength of the coronal magnetic field. The details of these procedures and their analysis codes have been described in previous studies and are publicly available, so we do not repeat them here \citep[e.g.,][]{2008SoPh..McIntosh,2009ApJ...Tomczyk,2015NatCo...Morton,2020Sci...369..694Y,2024Sci...386...76Y,zihao_yang_2024_13367069}.

As with most optically thin coronal observations, however, these observables are the result of multiple features contribution along the LOS. The derived physical quantities do not represent the plasma property at one single location and is not immediately clear how in a structured corona multiple features contribute along the LOS. A recent forward-modeling study using a MURaM simulation addressed this question \citep{2026ApJS..283....3Y}, but its LOS extent is relatively short and may not capture the level of LOS complexity typical of the corona. This motivates the forward-modeling approach adopted here. By synthesizing CoMP/UCoMP-like observables from a global MHD model with a much longer LOS integration length and more complex coronal structure, we compare the measured quantities with different model reference quantities and assess how LOS integration affects the interpretation of 2D coronal seismology measurements.

\subsection{Global MHD Model with Wave Driving}\label{sec:model}
We carried out a simulation of the global solar corona and solar wind with explicitly incorporated Alfv\'en waves using the Yin-Yang Magnetic Flux Eruption (Yin-Yang MFE) code \citep{2025ApJS..280...48L}. The MHD equations and the numerical algorithms used to solve the equations are described in detail in \citet{2025ApJS..280...48L}. The set up of the simulation is the same as the dipole solar wind simulation described in section 5 of \citet{2025ApJS..280...48L}, except the following specific changes.  For this simulation, we use a grid of $1020(r) \times 240 (\theta) \times 720 (\phi)$ for each Yin and Yang component, resulting in a uniform horizontal resolution of $0.375^{\circ}$. The radial range of the domain is $[R_{\odot}, 11.7 R_{\odot}]$ and the radial grid is stretched with a peak resolution of $4.46 \, {\rm Mm}$ near the lower boundary. We use an octupolar normal magnetic flux distribution on solar surface given by:
\begin{equation}
B_r (\theta) |_{R_{\odot}} = B_0 P_3(\cos\theta)
=\frac{B_0}{2}\left(5\cos^3\theta-3\cos\theta\right)
\label{eq:initflux}
\end{equation}
where $B_0 = 10$ G is the polar magnetic field strength.
Furthermore, we impose at the lower boundary two horizontal electric fields
to generate Alfv\'en waves and drive the emergence of a twisted flux rope
into one of the resulting helmet streamers.

\begin{figure}
    \centering
    \includegraphics[width=0.95\linewidth]{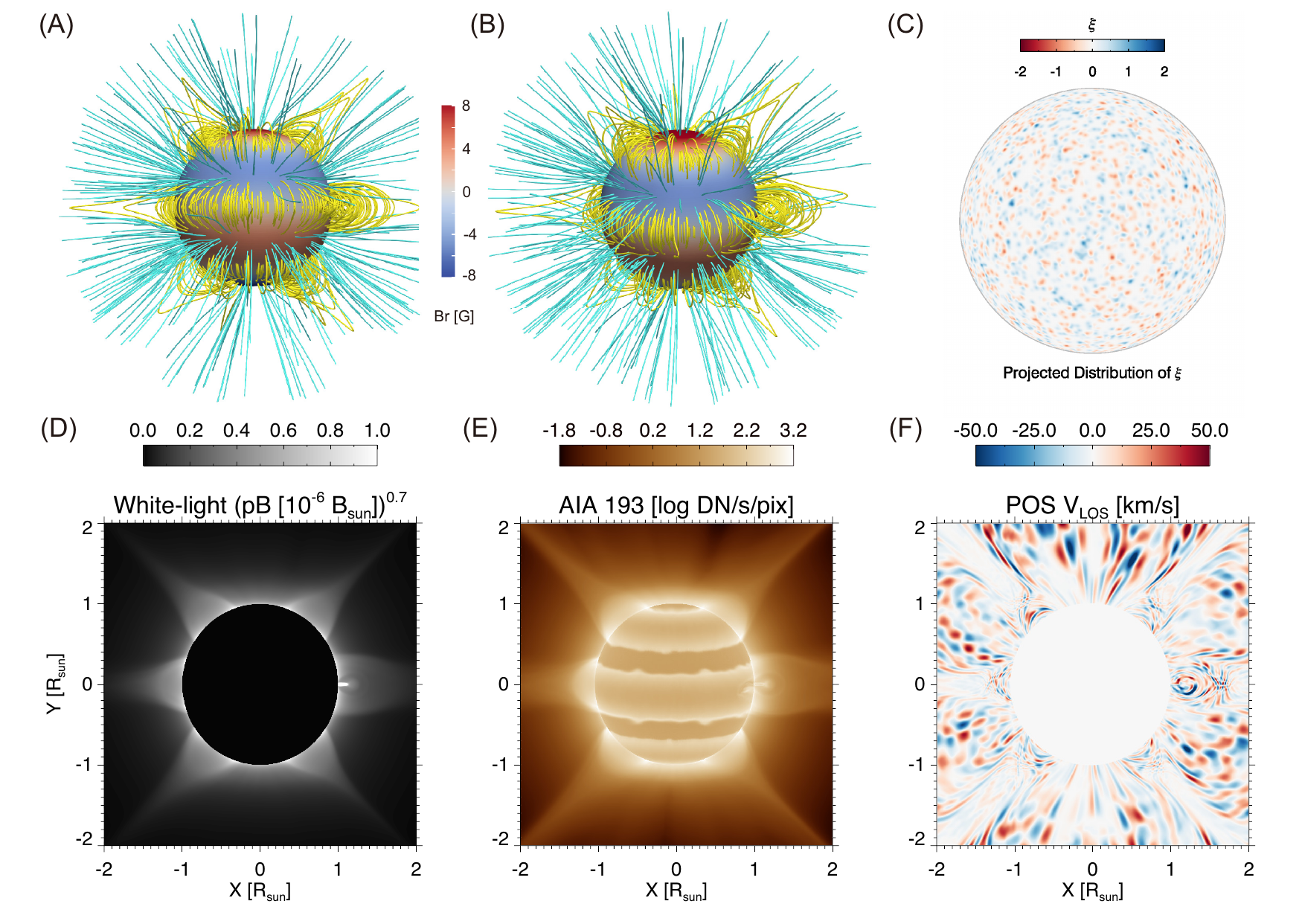}
    \caption{The quasi-steady global corona solution and the wave-driving boundary of the MHD model. (A) Open (cyan) and closed (yellow) magnetic field lines overlaid on the photospheric radial magnetic field. In this case, the magnetic field is viewed from the equator without a $B$-angle tilt. (B) Same as panel (A), but with the model viewed with a $10^{\circ}$ $B$-angle tilt. (C) A snapshot of the lower boundary cellular driving pattern $\xi$ in a spherical projection. The $\xi$ value is in unit of $2 \times 10^{15} \, {\rm cm}^2 \text{s}^{-1}$. An animation of this panel is available online. This animation shows the temporal evolution of the lower-boundary cellular driving pattern $\xi$ from time step 0 to 159, with an average cadence of $\sim35.68$ s, corresponding to a data duration of $\sim94.6$ min. The static panel shows a single snapshot at time step 0, while the animation shows the continuous appearance, evolution, and disappearance of the randomly distributed driving cells, illustrating the time-dependent boundary driving used to generate Alfv\'en waves in the simulation. (D) Snapshot of the synthetic white-light polarized brightness (PB) of the quasi-steady octupolar global MHD model viewed from the perspective of panel B. (E) Similar to panel D, but for AIA 193 \AA\ intensity. (F) Doppler velocity due to Alfv\'en wave fluctuations in the POS plane of (A) crossing the solar center. }
    \label{fig:model_fig}
\end{figure}

For generating the Alfv\'en waves, we impose the following random, horizontal
electric field $\bf{E}_{h}^{\rm{random}}$ \citep{2022ApJ...941...61F}:
\begin{equation}
c {\bf{E}}_{h}^{\rm{random}} = - \nabla_{h} ( \xi B_r ) ,
\label{eq:xi}
\end{equation}
where $c$ denotes the speed of light, the subscript $h$ denotes the
horizontal directions perpendicular to the normal/radial direction,
$B_r$ is the normal magnetic field on the lower boundary, and $\xi$ denotes
a time dependent cellular pattern that is composed of a superposition
of randomly placed cells of opposite sign values. Each of the cells has
a 2D Gaussian spatial profile with a width of 13.7 Mm and
a sinusoidal temporal profile with a period and life time of 5 min.
Figure \ref{fig:model_fig}(C) show a snapshot of the $\xi$ field.
An animation of the evolving $\xi$ field is available online.
${\bf{E}}_{h}^{\rm{random}}$ effectively drives random rotation
of the foot points of the $B_r$ flux concentrations, generating Alfv\'en
waves along the field lines.

For a period of time during the simulation we also impose
the emergence of a twisted magnetic torus (into the equatorial streamer) at the lower boundary by
imposing an additional horizontal electric field given by
equations (19) - (22) described in \citet{2017ApJ...844...26F}, with
the parameter $B_t a/R'$ for the axial field strength of the torus set to
68 G, and imposing the electric field for a duration of about 8.8 hours, after
which we stop the emergence and the emerged flux rope settles into a
quasi-equilibrium confined by the overlying corona streamer.

We numerically evolve the global coronal domain driven with the lower boundary
electric fields described above, until the corona reaches a quasi-steady
state.  The resulting quasi-steady global coronal solution is illustrated in
Figures \ref{fig:model_fig}(A)(B) and (D) to (F).
The octupolar magnetic field produces 6 coronal streamers and open coronal holes in
both the polar regions and mid latitude regions. In one of the streamers is embedded
a prominence cavity (see Figure \ref{fig:model_fig}(D)(E)))
corresponding to the twisted magnetic flux rope (Figure \ref{fig:model_fig}(A)(B))
with prominence condensation supported at its dips. Figure \ref{fig:model_fig}(F) shows a snapshot of the Doppler velocity in the POS of Figure \ref{fig:model_fig}(A) passing through the solar center, where ubiquitous Alfv\'en wave fluctuations can be seen.

\subsection{Construction of Synthetic UCoMP-like Observables}\label{sec:synth}
CoMP obtained spectroscopic information by sampling a coronal emission line at several discrete wavelength positions using a tunable narrow-band filter \citep{2008SoPh..247..411T}. UCoMP follows the same basic strategy. Its tunable birefringent filter is adjusted to shift the transmission profile in wavelength. In this way, the line profile is not measured continuously, but is instead sampled at a small number of tunings across the target line (e.g., the Fe \sc{xiii}\rm{} 1074.7 nm line). To mimic this observing process, we construct synthetic UCoMP-like observables by sampling the modeled line profile at three wavelength positions. In the main text, all results are shown for a model rotated by a $B$-angle of 10$^\circ$. We note that such a $B$-angle is not attainable in actual solar observations; here it is adopted simply to introduce greater line-of-sight structural complexity.

Before applying the three-point spectral sampling, we first synthesize the full Fe \textsc{xiii} line profiles from the physical quantities (i.e., density, temperature, velocity and magnetic field) in the global MHD model.

For each image pixel in the plane of sky (POS), we define a line of sight (LOS) extending from $-2\,R_\odot$ to $+2\,R_\odot$, where $R_\odot$ is the solar radius. Each LOS is uniformly sampled at 201 grid points. The field of view considered in this work extends to a maximum projected heliocentric distance of $2\,R_\odot$. The spatial pixel size is set to 3.48 Mm pixel$^{-1}$. At each sampled LOS position $i$, the local temperature, electron density, and LOS velocity are interpolated from the MHD model. Following Section 2.3 of \citet{2026ApJS..283....3Y}, we compute the local volume emissivity as
\begin{equation}
\epsilon_\text{i} \propto G(T_\text{i},n_{\rm e,i},h_\text{i})\,n_{{\rm e},\text{i}}^2,
\label{eq:eps_local}
\end{equation}
where $G(T,n_{\rm e},h)$ is the contribution function of the target line, obtained from a lookup table generated using CHIANTI version 11 \citep{1997A&AS..125..149D,2024ApJ...974...71D} as a function of temperature $T$, electron density $n_{\rm e}$, and coronal height $h$. Here both the electron collisional excitation and photo-excitation effects were included in the calculation of $G(T,n_{\rm e},h)$.

Assuming a Gaussian local line profile at each LOS position $\text{i}$, the spectral contribution from that LOS element is written as
\begin{equation}
\phi_\text{i}(\lambda)=
\frac{\epsilon_\text{i}}{\sqrt{2\pi}\,\sigma_{\lambda,\text{i}}}
\exp\left[-\frac{(\lambda-\lambda_\text{i})^2}{2\sigma_{\lambda,\text{i}}^2}\right],
\label{eq:localprof}
\end{equation}
where $\epsilon_\text{i}$ is the local volume emissivity, $\lambda_\text{i}=\lambda_{\rm rest}(1+v_{{\rm LOS},\text{i}}/c)$ is the Doppler-shifted line center, and $\sigma_{\lambda,\text{i}}=(\lambda_{\rm rest}/c)\sigma_{v,\text{i}}$ is the local spectral width. The velocity width is taken to be $\sigma_{v,\text{i}}=\sqrt{\sigma_{{\rm th},\text{i}}^2+\sigma_{\rm nt}^2}$, where $\sigma_{{\rm th},\text{i}}=\sqrt{k_B T_\text{i}/m_{\rm Fe}}$ is the thermal Gaussian width, $m_{\rm Fe}$ is the mass of the emitting Fe \textsc{xiii} ion. We adopt a value of $\sigma_{\rm nt}=15\ {\rm km\ s^{-1}}$ for the nonthermal Gaussian width. This value corresponds to a representative coronal nonthermal broadening scale reported in previous literature \citep[e.g.,][]{2009A&A...501L..15B,2011ApJ...727L..13B,2024ApJ...966..122Z}.

The LOS-integrated line profile at each POS pixel is then obtained by summing the contributions from all 201 LOS sampling points,
\begin{equation}
I(\lambda)\propto \sum_{\text{i}=0}^{200}\phi_\text{i}(\lambda)\,\Delta l,
\label{eq:los_profile}
\end{equation}
where $\Delta l$ is the uniform LOS sampling interval.

To mimic the UCoMP-like three-point spectral sampling, we convolve the LOS-integrated line profile with three instrumental transmission functions centered at the sampled wavelengths. For each target line, the three sampled intensities are defined as
\begin{equation}
I_j = \int I(\lambda)\,R_j(\lambda)\,{\rm d}\lambda, \qquad j=1,2,3,
\label{eq:three_point_sample}
\end{equation}
where $I(\lambda)$ is the LOS-integrated profile and $R_j(\lambda)$ is the transmission profile at the $j$th tuning position. Following our instrumental approximation, each transmission profile is assumed to be Gaussian, with a full width at half maximum (FWHM) of 0.138 nm \citep{2021AGUFMSH15G2089T}. In the calculation, the transmission functions are normalized to unit area.

For the Fe \textsc{xiii} 1074.7 nm line, the three sampling wavelengths are taken to be 1074.59 nm, 1074.70 nm, and 1074.81 nm; and for the Fe \textsc{xiii} 1079.8 nm line, the sampling wavelengths are 1079.69 nm, 1079.80 nm, and 1079.91 nm \citep{2024Sci...386...76Y}.

\begin{figure}
    \centering   
    \includegraphics[width=0.8\linewidth]{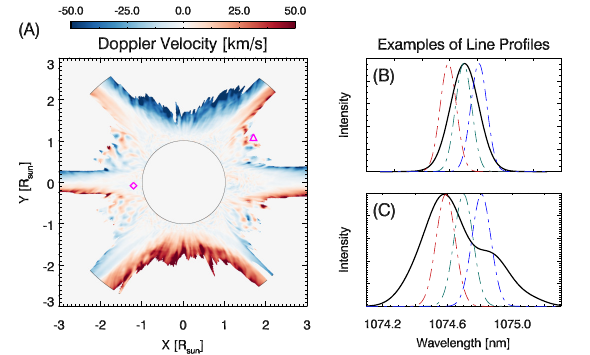}
\caption{Example of a Doppler-velocity map derived from three-point sampling of the Fe \sc{xiii}\rm{} 1074.7 nm line. (A) Doppler-velocity map. The magenta diamond marks a location where the three-point sampling yields a valid Doppler measurement, while the magenta triangle marks a location where the sampled intensities do not yield a physically meaningful Doppler measurement. In all figures in this work, only regions with valid Doppler measurements are included in the analysis. An animation of this panel is available. This animation shows the temporal evolution of the synthetic Doppler-velocity map from time step 0 to 159, with an average cadence of $\sim35.68$ s, corresponding to a data duration of $\sim94.6$ min. The static panel shows a single Doppler-velocity snapshot at time step 0, while the animation reveals the propagating Doppler-velocity fluctuations across the corona. These evolving wave signatures provide the temporal information used by the wave-tracking analysis to determine the wave propagation direction and phase speed. (B) LOS-integrated line profile (black solid curve) and the three transmission functions (colored dash-dotted curves) for the location marked  by magenta diamond in panel (A). (C) Same as panel (B), but for the the location marked by magenta triangle in panel (A).}
    \label{fig:dop_profile}
\end{figure}

For each pixel in the FOV, the three sampled intensities $(I_1,I_2,I_3)$ are computed using Eq. \ref{eq:three_point_sample}. We then apply an analytic Gaussian fitting method \citep{2013SoPh..288..637T} to derive the peak intensity and Doppler shift at each pixel. In the following analysis, we use the derived peak intensity as the synthetic intensity observable. Figure \ref{fig:dop_profile}(A) shows an example of the derived Doppler velocity map. In this work, we use 160 frames of Doppler velocity images, with an average cadence of $\sim$35.68 s. Fig. \ref{fig:dop_profile}(B) presents the corresponding LOS-integrated line profile and its three-point sampling at the location marked by the magenta diamond in panel (A). Some regions do not yield valid Doppler measurements because complex LOS velocity distributions distort the line profile and lead to nonphysical fitting results. An example is shown in Fig. \ref{fig:dop_profile}(C). Figure \ref{fig:density}(D) shows the Fe \textsc{xiii} 1074.7 nm intensity derived from the analytic Gaussian fit.

\section{Comparison of Measurements with Model Reference Quantities}

Using the synthetic line-intensity and Doppler-velocity observables described above, we apply the same 2D coronal seismology procedures used in previous studies to derive the density, wave propagation direction, phase speed, and magnetic field. The technical details of these diagnostics have been presented in previous literature \citep[e.g.,][]{2020Sci...369..694Y,2020ScChE..63.2357Y,2024ApJ...963...69B,2024Sci...386...76Y,2026ApJS..283....3Y} and are not repeated here. Instead, we compare the measured quantities with different model reference quantities to determine what physical quantities are actually measured by 2D coronal seismology. In particular, we examine whether the measurements are best described by simple LOS emissivity-weighted quantities as suggested in previous literature \citep[e.g.,][]{2024Sci...386...76Y, 2026ApJS..283....3Y} or whether additional weighting effects, such as wave amplitude dependence, become important.

\subsection{Density Diagnostics}\label{sec:density}

\begin{figure}
    \centering   
    \includegraphics[width=0.7\linewidth]{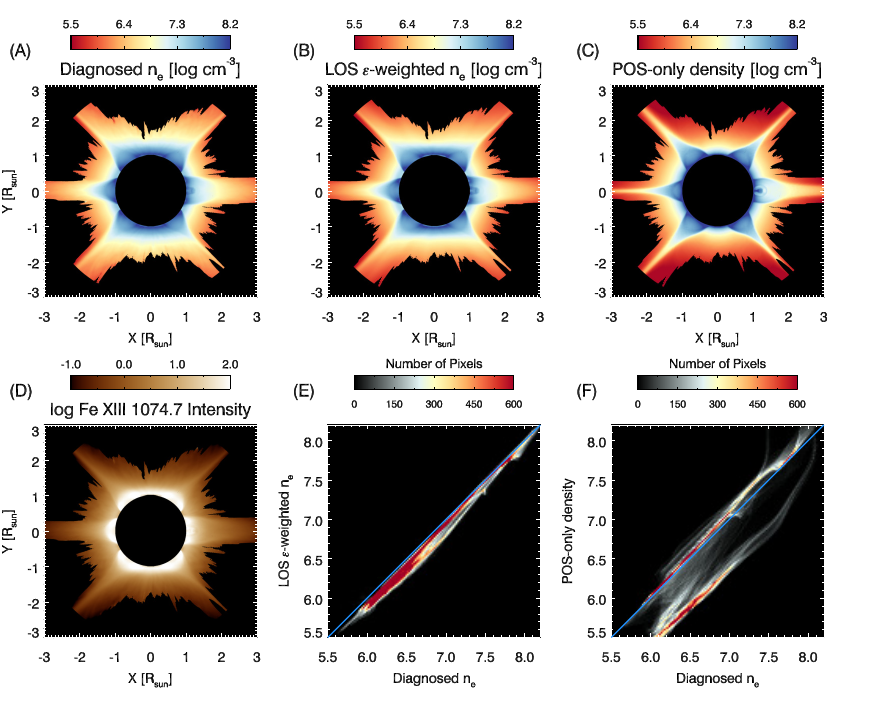}
\caption{Density diagnostic results. (A) Diagnosed electron-density map derived from the intensity ratio of the synthetic Fe \sc{xiii}\rm{} lines. (B) LOS emissivity-weighted average electron density calculated directly from the model. (C) The electron density from the model on the POS slice crossing the solar center. (D) Intensity map of the Fe \textsc{xiii} 1074.7 nm line derived from the three-point sampling and analytic Gaussian fit. (E) Two-dimensional histogram comparing the diagnosed electron density with the LOS emissivity-weighted average density from the model. The blue line marks the one-to-one relation. (F) Similar to (E), but comparing the diagnosed one with the POS-only density.}
    \label{fig:density}
\end{figure}

The intensity ratio of the two Fe \textsc{xiii} infrared lines has been widely used to diagnose electron density in astrophysical plasmas \citep[e.g.,][]{2021ApJ...Dudik,2024ApJ...966..122Z,2026ApJ...FeXIIINova}. Figure~\ref{fig:density}(A) shows the electron-density map diagnosed from the synthetic observables. We use CHIANTI version 11 to calculate the theoretical relationship between the Fe \textsc{xiii} line ratio and electron density, including both electron collisional excitation and photo-excitation.

Because the density diagnostic is based on LOS-integrated line emission in an optically thin corona, the key question is which model quantity the diagnosed density most closely represents. Emissivity-weighted quantities are commonly used as reference values when interpreting physical parameters derived from LOS-integrated observations. For example, temperatures diagnosed from optically thin observations are often interpreted as LOS emissivity-weighted temperatures \citep[e.g.,][]{2020ApJ...Gilly}; \citet{2024ApJ...faraday} described the magnetic field inferred from Faraday rotation as an LOS emissivity-weighted average; and in \citet{2021ApJ...yajie}, the density, temperature, and magnetic field diagnosed from the magnetic-field-induced-transition method were all interpreted as emissivity-weighted quantities. We therefore compare the diagnosed density with the LOS emissivity-weighted electron density, which is expected to approximate the effective density sampled by the diagnostic. 

We define the model reference density as:
\begin{equation}\label{eq:weighted_dens}
\left<n_\text{e}\right>=\frac{\int \epsilon_\text{i} n_{\text{e},\text{i}}\,\text{d}l}{\int \epsilon_\text{i}\,\text{d}l},
\end{equation}
where $n_{\text{e},\text{i}}$ is the electron density at LOS location $\text{i}$, as obtained from the MHD model. As in the line-profile synthesis described in Sect.~\ref{sec:synth}, the LOS integration is performed over 201 uniformly sampled locations between $-2\,R_\odot$ and $+2\,R_\odot$.

Figure~\ref{fig:density}(B) shows the LOS emissivity-weighted density calculated directly from the model. Figure~\ref{fig:density}(E) presents a 2D histogram comparing the diagnosed density from the synthetic observables with the emissivity-weighted density from the model. The close correspondence between the two quantities indicates that the density inferred from the line ratio is well represented by a simple LOS emissivity-weighted average. We also compared the diagnosed density with the density on the POS slice crossing the solar center in Figure \ref{fig:density}(C) and (F). Larger discrepancies are evident in this comparison. Additional examples for model rotations of $B$-angle=$0^\circ$ and $B$-angle=$20^\circ$ (where positive $B$-angle corresponds to a tilt toward the observer) reveal the same overall behavior and are shown in Appendix \ref{ap:add}.

\subsection{Wave Propagation Direction}\label{sec:angle}

We next evaluate the wave propagation direction and phase speed derived from the synthetic Doppler velocity observables. 

In 2D coronal seismology, these two quantities provide the key link to the magnetic field: the wave propagation direction traces the POS magnetic-field orientation because transverse waves propagate along magnetic field lines, while the phase speed, together with the diagnosed density, is used to infer the magnetic-field strength. Both quantities are obtained through the wave-tracking analysis of the Doppler velocity time series. The local wave propagation direction (or wave angle) is inferred from the high-coherence region in the coherence map constructed from the cross-correlation of neighboring Doppler signals. Using the calculated wave propagation direction, a wave path of fixed length is defined to construct the Doppler velocity time--distance diagram, from which the phase speed is derived. Here we do not repeat the full details of the wave-tracking procedure and instead refer the reader to previous studies for technical details \citep[][]{2020ScChE..63.2357Y,2024Sci...386...76Y,2026ApJS..283....3Y}.

Figure~\ref{fig:vph}(A) shows the map of the measured wave propagation direction. The local radial direction is taken as the reference direction. Angles measured clockwise from the local radial are defined as negative, while those measured counterclockwise are defined as positive.

Following previous work \citep[e.g.,][]{2024Sci...386...76Y,2026ApJS..283....3Y}, we use the LOS emissivity-weighted POS magnetic-field direction from the model as the reference quantity for comparison with the measured wave propagation direction. This reference quantity may be written as
\begin{equation}
    \left<\Psi\right>=\frac{\int \epsilon_\text{i}\,\Psi_\text{i}\,\text{d}l}{\int \epsilon_\text{i}\,\text{d}l},
\end{equation}
where $\Psi_\text{i}$ is the local POS magnetic-field direction at LOS location $\text{i}$. However, because the POS magnetic-field direction is an axial quantity rather than a vector angle, $\Psi$ and $\Psi+180^\circ$ represent the same physical orientation. Therefore, the LOS-weighted reference direction cannot be obtained by directly averaging the local angles along the LOS, as this would lead to artificial cancellation near the $\pm90^\circ$ boundary. Instead, for each LOS position we first compute the local POS magnetic-field angle and then perform the LOS weighting in double-angle space by averaging $\cos(2\Psi)$ and $\sin(2\Psi)$ with emissivity. The mean POS magnetic-field direction is then recovered by dividing the resulting angle by two.

Figure~\ref{fig:vph}(B) shows the emissivity-weighted POS magnetic-field direction calculated from the model. The comparison in Figure~\ref{fig:vph}(C) shows a close correspondence between this reference quantity and the wave-tracking result. This indicates that, similar to the density diagnostic, the measured wave propagation direction is well represented by an LOS emissivity-weighted average and therefore provides a reliable proxy for the LOS-weighted POS magnetic-field direction. The two additional viewing-angle cases shown in Appendix~\ref{ap:add} also demonstrate the same result.

\subsection{Phase Speed}\label{sec:vph}

Despite the much longer LOS extent and the substantially more complex coronal structure in the present global MHD model, the density and wave propagation direction diagnostics remain broadly consistent with the behavior found previously in the small-domain MURaM simulation of \cite{2026ApJS..283....3Y} with a much shorter LOS. However, the phase-speed diagnostic is more complicated and nonlinear, making its interpretation less straightforward. 

The measured phase speed is not determined by a simple algebraic combination of LOS contributions. Instead, it is derived from the propagation pattern of the LOS-integrated Doppler velocity along the tracked wave path. Moreover, what is tracked in practice is not the Doppler velocity itself, but the propagating redshift and blueshift oscillation pattern in the Doppler signal. Therefore, the phase speed measurement is controlled primarily by the LOS components that produce the clearest oscillatory propagation signature from one position to the next. In particular, LOS regions with larger velocity amplitudes can contribute more strongly to the measured phase speed, because their Doppler oscillations leave a stronger and more traceable propagation pattern in the time-distance diagram. As a result, the measured phase speed can depend on velocity amplitude in addition to emissivity. 

\begin{figure}
    \centering   
    \includegraphics[width=0.9\linewidth]{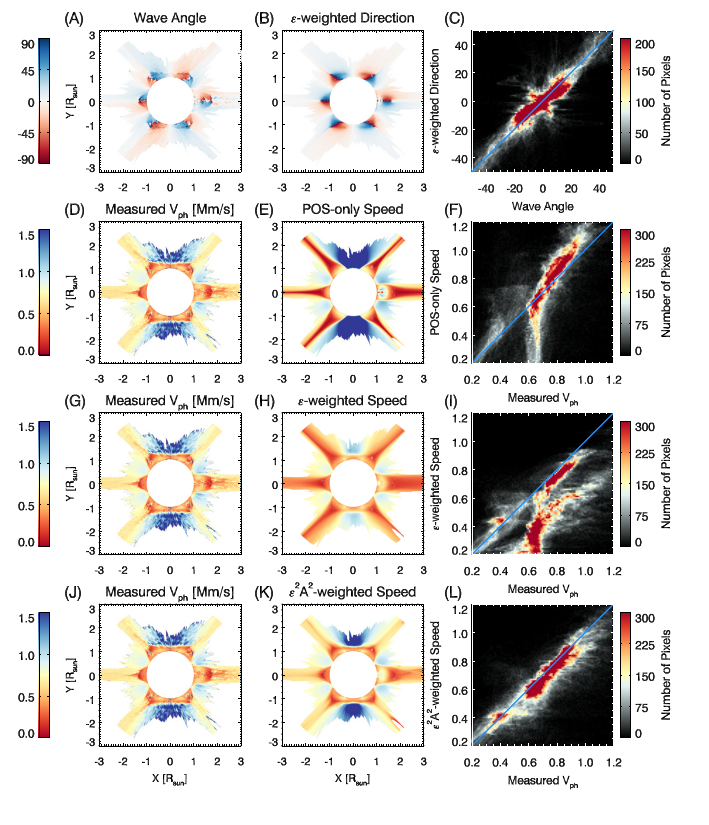}
\caption{Wave propagation direction and phase-speed measurements. (A) Measured wave propagation direction (wave angle) derived from the wave-tracking procedure applied to the synthetic Doppler-velocity data. (B) LOS emissivity-weighted average POS magnetic-field direction calculated from the model. (C) Two-dimensional histogram comparing the measured wave angle with the reference magnetic-field direction. The blue line marks the one-to-one relation. (D) Measured phase speed derived from the wave-tracking procedure. (E) The propagation speed at the POS slice crossing the solar center. To account for the contribution from background flows, this reference speed includes both the wave phase speed and the outflow speed. (F) Same as panel (C), but comparing the measured phase speed in panel (D) with the POS-only reference speed in panel (E). (G) Same as panel (D), shown again for comparison with the emissivity-weighting reference speed. (H) LOS emissivity-weighted average propagation speed calculated from the model. (I) Same as panel (C), but comparing the measured phase speed in panel (G) with the emissivity-weighted reference speed in panel (H). (J) Same as panel (D), shown again for comparison with the amplitude-dependent reference speed. (K) Propagation speed calculated from the model using the $\epsilon^2A^2$ weighting. (L) Same as panel (C), but comparing the measured phase speed in panel (J) with the $\epsilon^2A^2$-weighted reference speed in panel (K).}
    \label{fig:vph}
\end{figure}

Figure~\ref{fig:vph}(D) shows the measured phase speed derived from the synthetic Doppler velocity data. 
In the previous forward-modeling study based on a short-LOS MHD simulation, the measured phase speed was found to agree well with the LOS emissivity-weighted phase speed \citep{2026ApJS..283....3Y}. For comparison, we also calculate a LOS-weighted reference propagation speed for the present global model. We note, however, that the background outflow in the present model is non-negligible and can therefore affect the phase speed measured from wave tracking. To account for this effect, we define the emissivity-weighted effective propagation speed as
\begin{equation}\label{eq:vph_weight}
    \left<v_\text{POS}\right>=\frac{\int \omega_\text{i}\,[v_\text{ph,POS}(\text{i})+v_\text{flow,POS}(\text{i})]\,\mathrm{d}l}{\int \omega_\text{i}\,\mathrm{d}l},
\end{equation}
where $v_\text{ph,POS}(\text{i})$ is the local POS phase speed at LOS position $\text{i}$, calculated from the POS magnetic-field strength and density in the model, $v_\text{flow,POS}(\text{i})$ is the local POS outflow speed, and $\omega_\text{i}$ is the adopted weight.

Before comparing the measured propagation speed with the weighted propagation speed, we first compared it with the effective propagation speed on the POS slice, which includes both the phase speed and the outflow speed (Figure \ref{fig:vph}(E)). This comparison shows clear discrepancies, particularly in the plasma-sheet and polar regions.

Figure~\ref{fig:vph}(H) shows the effective propagation speed weighted by emissivity alone ($\omega=\epsilon$). Clear discrepancies are present between panels (D) and (H). The 2D histogram in panel (I) further shows that the measured phase speed is well approximated by the emissivity-weighted speed only over part of the FOV. This behavior is in clear contrast to the previous short-LOS forward-modeling result of \cite{2026ApJS..283....3Y}.

As discussed above, velocity amplitude can become a non-negligible additional weighting factor in the present model. To construct a reference quantity that better matches the measurement, we further include the LOS velocity fluctuation amplitude in the weighting. The local LOS-projected velocity amplitude is estimated from
\begin{equation}\label{eq:amp}
    A_\text{i}=\sqrt{
\left< v_{\mathrm{LOS}}(\text{i},t)^2 \right>_t-\left< v_{\mathrm{LOS}}(\text{i},t) \right>_t^2},
\end{equation}
where $v_{\mathrm{LOS}}(\text{i},t)$ is the local LOS velocity time series at LOS position $\text{i}$. Equation~\ref{eq:amp} gives the temporal standard deviation of the local LOS velocity and is used here as a measure of the velocity amplitude projected along the LOS.

 Following Equation~\ref{eq:vph_weight}, we further calculate the effective propagation speeds using $\omega_1=\epsilon A$ and $\omega_2=\epsilon^2 A^2$, respectively. The reference quantity weighted by $\omega_2=\epsilon^2 A^2$ is shown in Figure~\ref{fig:vph}(K), and panel (L) compares this reference quantity with the measured phase speed. By contrast, the result obtained with $\omega_1=\epsilon A$ still shows noticeable discrepancies from the measurement and is therefore not displayed here. A more detailed discussion on why velocity amplitude affects the phase-speed measurement, and why $\epsilon^2A^2$ provides a better effective weighting, is provided in Appendix \ref{app:vph_weight}. 

The $\epsilon^2 A^2$-weighted reference shows a much closer correspondence to the measured phase speed than the emissivity-only case. This comparison indicates that the measured phase speed is influenced not only by emissivity, but also by the distribution of velocity amplitude along the LOS. We do not interpret the $\epsilon^2 A^2$ weighting as a unique analytic result. Rather, it should be regarded as an effective empirical weighting that best matches the forward-modeled measurements in the present model. These results demonstrate that amplitude weighting can become important for interpreting wave-tracking measurements when the LOS structure is sufficiently complex. Additional viewing-angle cases shown in Appendix~\ref{ap:add} exhibit the same overall behavior. To demonstrate the role of amplitude weighting in the measurement of phase speed, we constructed an idealized toy model (see Appendix \ref{ap:toymodel}).

\subsection{Magnetic Field Strength}

\begin{figure}
    \centering   
    \includegraphics[width=0.99\linewidth]{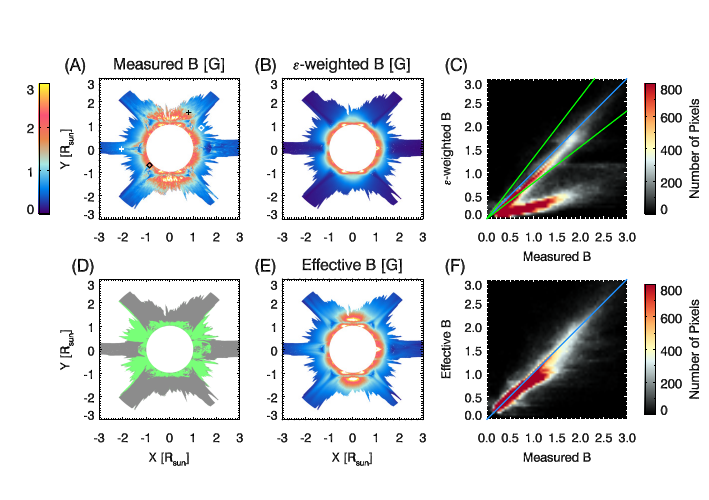}
\caption{Magnetic-field measurements. (A) Coronal magnetic-field strength inferred from the synthetic observables using the 2D coronal seismology technique. The black and white diamonds and crosses mark four representative locations selected for further analysis of the LOS distribution of velocity amplitude, as shown in Figure \ref{fig:weights}. (B) LOS emissivity-weighted POS magnetic-field strength calculated from the model. (C) Two-dimensional histogram comparing the measured magnetic-field strength with the LOS emissivity-weighted magnetic-field strength. The blue line marks the one-to-one relation. The two green lines indicate the range within which the two magnetic-field estimates differ by less than a factor of 1.3, i.e., $1/1.3 \le \left<B_\text{POS}\right>/B_{\rm measured} \le 1.3$. Pixels within this range correspond to the regions highlighted by the green mask in panel (D). (D) The green masked regions indicate pixels satisfying this agreement criterion, while the grey masked regions indicate the remaining valid pixels. (E) Effective magnetic-field strength calculated from the model using Equation~\ref{eq:beff}. (F) Same as panel (C), but comparing the measured magnetic-field strength in panel (D) with the effective magnetic-field strength in panel (E).}
    \label{fig:bfield}
\end{figure}

Following Equation~\ref{eq:kinkeq}, the magnetic-field strength can be inferred from the measured density and phase speed obtained in Sections~\ref{sec:density} and \ref{sec:vph}, and the results are shown in Figure~\ref{fig:bfield}(A).

As discussed above, for optically thin astrophysical observations, emissivity-weighting provides a natural baseline for accounting for LOS integration effects \citep[e.g.,][]{2020ApJ...Gilly,2021ApJ...Dudik,2021ApJ...yajie,2024ApJ...faraday}. This assumption was also adopted in the previous UCoMP measurement study \citep{2024Sci...386...76Y} and in earlier forward-modeling validation works \citep[e.g.,][]{2025RAA....Gao,2026ApJS..283....3Y}. Because 2D coronal seismology traces wave signatures in the POS, these previous studies used the LOS-weighted POS magnetic-field strength as the reference quantity. As a baseline comparison, we therefore calculate the LOS emissivity-weighted POS magnetic-field strength from the model,
\begin{equation}
    \left<B_\text{POS}\right>=\frac{\int \epsilon_\text{i} B_\text{POS}(\text{i})\,\mathrm{d}l}{\int \epsilon_\text{i}\,\mathrm{d}l}.
\end{equation}
The result is shown in Figure~\ref{fig:bfield}(B). Large discrepancies are evident between this reference quantity and the measured magnetic-field strength, especially in the polar regions and the plasma-sheet cusp regions. The 2D histogram in panel (C) further reveals two main branches: one corresponding to relatively good agreement between the emissivity-weighted magnetic field and the measured field, and the other to substantial disagreement. We define relatively close agreement as the two magnetic-field estimates differing by less than a factor of 1.3, i.e., $1/1.3 \le \left<B_\text{POS}\right>/B_{\rm measured} \le 1.3$, corresponding to the region between the two green lines in Figure~\ref{fig:bfield}(C). The locations of these pixels are highlighted with a green mask in Figure~\ref{fig:bfield}(D), while the remaining  pixels with disagreement are shown in gray.

To construct a model reference quantity that is more closely matched to the measurement, we return to Equation~\ref{eq:kinkeq}, which forms the basis of the magnetic-field inference in 2D coronal seismology. The measured magnetic field is not directly comparable to an emissivity-weighted magnetic field, because it is derived from a nonlinear combination of the measured density and phase speed. Moreover, as shown in the previous sections, the density and phase speed are themselves sensitive to different LOS weighting factors.

A practical way to construct an effective reference magnetic field is therefore to combine the $\epsilon$-weighted density (Equation~\ref{eq:weighted_dens} and Figure~\ref{fig:density}(B)) with the $\epsilon^2 A^2$-weighted propagation speed (Equation~\ref{eq:vph_weight} and Figure~\ref{fig:vph}(K)), and then apply Equation~\ref{eq:kinkeq} to obtain
\begin{equation}\label{eq:beff}
    B_\text{eff}=\left<v_\text{POS}\right>_{\omega=\epsilon^2 A^2}\sqrt{\mu_0\left<\rho\right>_{\omega=\epsilon}}.
\end{equation}

Figure~\ref{fig:bfield}(E) and (F) show the resulting effective magnetic-field strength and the corresponding 2D histogram comparing it with the measured field in Fig. \ref{fig:bfield}(A). The agreement is now much improved, and the two maps share a high degree of similarity. This result suggests that simple emissivity weighting alone can still provide a reasonable approximation in some regions, but that the more appropriate reference quantity for the measured magnetic field is likely the effective field constructed from Equation~\ref{eq:kinkeq}, with the phase speed weighted by both emissivity and velocity amplitude. Additional viewing-angle cases in Appendix~\ref{ap:add} show the same overall behavior. We discuss the implications further in the next section.

\section{Discussion}
\subsection{The Role of Velocity Amplitude in LOS Weighting}
As shown in Section~\ref{sec:density}, the density diagnosed from the Fe \textsc{xiii} line ratio is well represented by an LOS emissivity-weighted average density. By contrast, Figure~\ref{fig:vph} shows that the measured phase speed is controlled not only by the emissivity distribution along the LOS, but also by the distribution of velocity amplitude among different LOS structures. As a result, the inferred magnetic field is likewise affected by these different weighting factors. In this section, we investigate how the LOS distribution of velocity amplitude influences the magnetic-field measurement and under what conditions its role becomes non-negligible.

The green masked regions in Figure~\ref{fig:bfield}(D) indicate the part of the FOV where the measured magnetic field can still be reasonably approximated by the LOS emissivity-weighted average magnetic field. By contrast, the surrounding regions masked by gray color correspond to locations where velocity amplitude plays a more important role in the LOS weighting. To illustrate how this behavior depends on the LOS distribution of coronal properties, we select four representative locations in the FOV. Two of them, marked by the black and white diamonds in Figure~\ref{fig:bfield}(A), are cases where emissivity-weighting remains a good approximation to the measured magnetic field. The other two, marked by the black and white crosses, are representative cases where velocity amplitude becomes a non-negligible weighting factor.

\begin{figure}
    \centering   
    \includegraphics[width=0.76\linewidth]{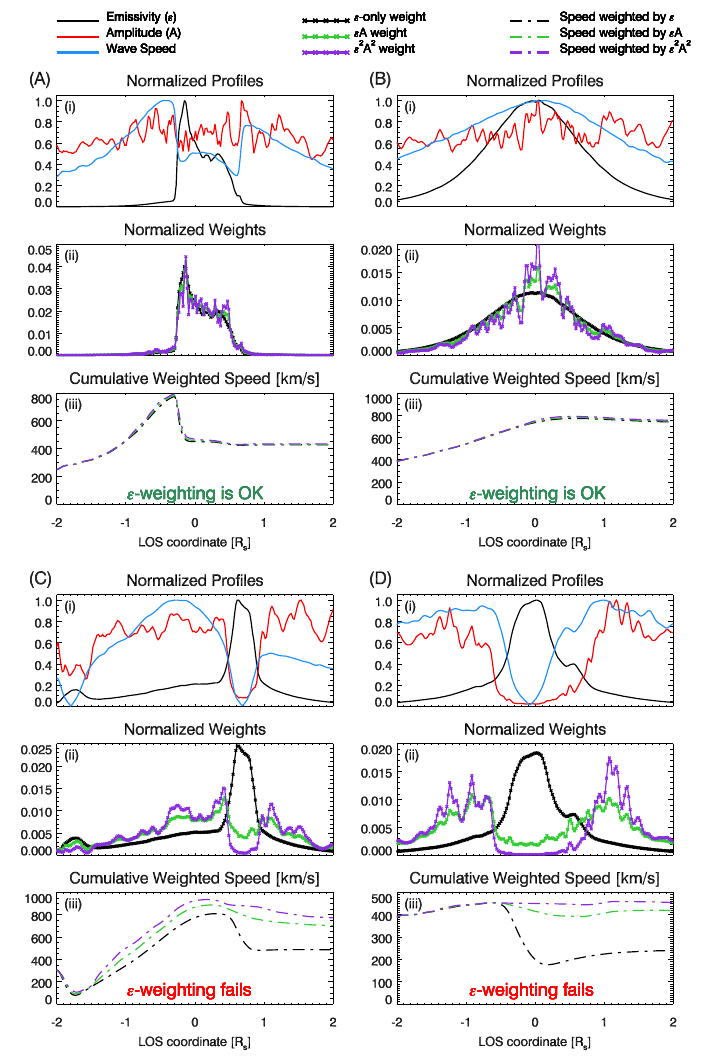}
    \caption{LOS distributions of velocity amplitude and their impact on LOS weighting. The figure consists of four panels, each containing three subpanels. (A) Results from the black diamond marked in Figure \ref{fig:bfield}(A). (B) Results from the white diamond marked in Figure \ref{fig:bfield}(A). (C) Results from the black cross marked in Figure \ref{fig:bfield}(A). (D) Results from the white cross marked in Figure \ref{fig:bfield}(A). Subpanel (i): Normalized LOS profiles of emissivity (black), velocity amplitude (red), and wave speed (blue). Subpanel (ii): Normalized LOS weighting profiles for $\epsilon$ (black crosses), $\epsilon A$ (green crosses), and $\epsilon^2A^2$ (purple crosses). Subpanel (iii): Cumulative LOS-weighted speed weighted by $\epsilon$ (black dot-dashed curve), $\epsilon A$ (green dot-dashed curve), and $\epsilon^2A^2$ (purple dot-dashed curve).}
    \label{fig:weights}
\end{figure}

Figure~\ref{fig:weights} summarizes the LOS distributions of different coronal properties, the corresponding LOS weights, and the cumulative weighted speed for four representative locations. Panels (A) and (B) correspond to the black and white diamonds in Figure~\ref{fig:bfield}(A), respectively, whereas panels (C) and (D) correspond to the black and white crosses.

Each panel in Figure~\ref{fig:weights} contains three subpanels. Subpanel (i) shows the normalized LOS profiles of emissivity ($\epsilon$, black solid line), velocity amplitude ($A$, red solid line) calculated from Equation~\ref{eq:amp}, and wave speed ($v_\text{POS}=v_\text{ph,POS}+v_\text{flow,POS}$, blue solid line). Subpanel (ii) shows the normalized LOS weighting functions for $\epsilon$ only (black crosses), $\epsilon A$ (green crosses), and $\epsilon^2 A^2$ (purple crosses). Subpanel (iii) shows the cumulative weighted speed along the LOS, where the black, green, and purple dash-dotted lines correspond to weighting by $\epsilon$, $\epsilon A$, and $\epsilon^2 A^2$, respectively.

We first examine the two examples from regions where emissivity-weighting still provides a good approximation (Fig. \ref{fig:weights}(A) and (B)). In both cases, the velocity amplitude (red solid line) varies along the LOS, but does not exhibit strong local enhancements or depressions. Its variation remains mostly within one order of magnitude. As a result, the three weighting functions shown in subpanel (ii) have very similar distributions along the LOS, indicating that they place comparable emphasis on the same structures. The cumulative weighted speeds in subpanel (iii) likewise remain very similar, showing that the inclusion of velocity amplitude has little effect on the LOS-integrated result. These two examples suggest that when the LOS variation of velocity amplitude is modest, roughly within one order of magnitude, its contribution to the weighting is likely to be small. In this regime, emissivity-weighting remains a good approximation for the quantities measured by 2D coronal seismology. More examples are shown in the Appendix \ref{sec:app_weight}.

The behavior is very different in the other two examples, where emissivity-weighting fails (Fig. \ref{fig:weights}(C) and (D)). In both cases, the velocity amplitude shows sharp drops and enhancements along the LOS, occurring near locations where the emissivity and phase speed also change strongly. As described in the model overview, the global MHD model contains several dense plasma-sheet cusp regions. These structures have enhanced emissivity but very weak magnetic field, and therefore low phase speed and weak velocity amplitude. The two LOS examples in Figure~\ref{fig:weights}(C) and (D) are cases in which the LOS crosses such plasma-sheet regions. Because the phase-speed measurement is determined by tracking the pattern of Doppler velocity fluctuations, LOS regions with larger velocity amplitudes exert a stronger influence on the measured phase speed. When a dense, highly emissive plasma-sheet region has only weak velocity fluctuations, its contribution to the wave-tracking measurement can be outweighed by neighboring structures with larger amplitudes. In that case, a simple emissivity-weighting overrepresents the plasma-sheet contribution. This is clearly reflected in subpanels (ii) and (iii) of Figure~\ref{fig:weights}(C) and (D), where the $\epsilon$-only weighting differs substantially from the amplitude-dependent weightings. This explains why the measured phase speed and magnetic field deviate strongly from the quantities weighted by emissivity alone.

From the discussion above, it is important to understand the LOS distribution of velocity amplitudes when interpreting the quantities measured from 2D coronal seismology. However, this information cannot be directly obtained from observations. Although the LOS distribution of wave amplitude cannot be directly recovered from a single-view observation, several observational signatures may help identify LOSs for which amplitude-dependent effects are more likely to be important. Complex LOS velocity distributions may produce asymmetric or non-Gaussian line profiles, enhanced line widths, or poor single-component spectral fits; indeed, some complex LOSs in our forward modeling produce distorted line profiles for which the three-point Doppler diagnostic fails (e.g., Section \ref{sec:synth}). Complementary EUV, coronal-line intensity, and white-light observations may also help identify LOSs that intersect multiple or strongly structured coronal features. Additional constraints on the coronal structures intersected by the LOS may be obtained by combining 2D coronal seismology with coronal models, tomography, or stereoscopic observations from multiple instruments. These approaches do not uniquely determine the LOS distribution of wave amplitudes, but they may provide practical ways to flag measurements that are potentially more susceptible to amplitude-dependent LOS effects.

Another relevant consideration is the range of wave amplitudes typically present in the corona. Although no previous study has examined how velocity amplitudes are distributed in different regions based on CoMP/UCoMP observations, other observed kink-like transverse oscillations and waves may share similar amplitude properties, including velocity amplitude, with the transverse waves observed by CoMP/UCoMP. A form of decayless kink oscillation was independently identified in imaging and spectroscopic observations \citep{2012ApJ...Wang,2012ApJ...tian}, and its properties, including velocity amplitude, have been studied extensively. Because these oscillations also exhibit kink-like behavior, they may provide a useful comparison with the kink-like waves observed by CoMP/UCoMP. Extensive work has been carried out to characterize the distribution of velocity amplitudes of such kink oscillations in different regions. \cite{2016A&A...Nakariakov} found that the velocity amplitudes of undamped oscillations are mostly in the range of 2--8 km s$^{-1}$ in coronal loops of varying lengths. A more recent statistical study by \cite{2024A&A...Lim} measured the velocity amplitudes of these oscillations in different active regions and quiet-Sun regions over three years of observations. They found that, although the velocity amplitude can vary substantially, the majority of the values fall within the range of 1--10 km s$^{-1}$. In another study, \citet{gao2026high} detected the propagating kink wave signals in polar plumes and found that for waves near 3.5 mHz, their velocity amplitudes are generally between 1 and 10 km s$^{-1}$. These studies suggest that the velocity amplitudes of these kink-like oscillations/waves typically vary within one order of magnitude. If the same is true for the kink-like waves observed by CoMP/UCoMP, then, as long as the LOS does not pass through regions with strong property variations (e.g., plasma-sheet regions), the impact of velocity amplitude variations on 2D coronal seismology diagnostics may be negligible, and emissivity weighting should still provide a good estimate.

The results in Figure \ref{fig:weights}(C)(D) suggest that if the LOS crosses coronal regions with strong variations in physical properties, where both the velocity amplitude and emissivity change rapidly along the LOS, measurements from 2D coronal seismology will be significantly affected by the velocity-amplitude distribution. In such cases, the amplitude must be taken into account when interpreting LOS integration effects. Such regions include plasma sheets and streamer cusps. In the global MHD model used in this work, a conical plasma-sheet structure extends outward from each polar region, and a plasma-sheet belt also surrounds the equatorial region. Under this model setup, the LOS therefore has a high probability of crossing plasma-sheet regions. For such LOS, the measured phase speed and magnetic field can depend strongly on the velocity amplitude.

We note that the quantitative form of the amplitude-dependent weighting obtained here may be model dependent. The $\epsilon^2A^2$ weighting should be regarded as an effective empirical weighting for the present simulation rather than a universal prescription for 2D coronal seismology. Current global coronal MHD models have limited spatial resolution and cannot fully capture the fine-scale dynamics and structural complexity of the real corona. Recent coronal observations have revealed substantial fine-scale structuring in the extended corona \citep[e.g.,][]{2026A&A...710A.138A}, suggesting that the LOS complexity in the real corona may exceed that represented in the present simulation in some cases. Such unresolved structures may produce different distributions of emissivity, velocity amplitude, and phase speed along the LOS and therefore modify the effective weighting. The dependence of the LOS weighting on wave period has not been systematically examined in the present model. Nevertheless, we do not expect modest changes in wave period to qualitatively alter the weighting mechanism. In our previous forward-modeling study, the MURaM wave field exhibited power over a broad range of frequencies, and the inferred phase speeds were insensitive to the precise filtering frequency when appropriate wave-tracking parameters were adopted \citep[][]{2026ApJS..283....3Y}. The more general result of the present analysis is that velocity amplitude can contribute to the LOS weighting of the measured phase speed, while the quantitative form of this weighting requires further investigation with different models and observations.

Observationally, global large-scale plasma sheets are usually located at heights of at least 1.5--2 $R_\odot$. For example, the transition from the closed arcade to the narrow streamer stalk occurs at the streamer cusp, and such regions have been reported to lie at $\sim 2\,R_\sun$ \citep[e.g.,][]{2024SoPh..299...WangYM,2020ApJ...893...Morgan}. In contrast, the CoMP/UCoMP FOV for coronal magnetic-field measurements extends to at most 1.6 $R_\sun$ \citep{2024Sci...386...76Y}. Therefore, the particularly strong amplitude-weighting effects associated with the large-scale plasma-sheet structures in the present model may be less prevalent in typical CoMP/UCoMP measurements. Nevertheless, real LOS measurements within the CoMP/UCoMP FOV can still pass through other complex and highly structured coronal regions. Smaller-scale plasma sheets, for example, can occur at various locations and heights in the corona \citep[e.g.,][]{2018ApJ...ChengXin,2019ApJ...French,2024ApJ...973...CS}, and other forms of fine-scale coronal structuring may similarly produce substantial variations in emissivity and wave amplitude along the LOS. Therefore, the plasma-sheet regions in the present model should be regarded as representative examples of conditions under which amplitude-dependent weighting becomes important, rather than as the only structures capable of producing such effects. The importance of these more general forms of LOS complexity for CoMP/UCoMP measurements remains to be quantitatively assessed. Figures \ref{fig:bfield}(A) and (D) also show that for the low coronal regions and open-field regions without plasma-sheet structures, the measured magnetic-field strength remains relatively close to the LOS emissivity-weighted magnetic-field strength.

\subsection{The Impact of Background Flow}

As discussed in Section \ref{sec:vph}, the outflow speed is non-negligible in the present MHD model, especially in open-field regions. This background flow contributes to the apparent phase speed and therefore affects the seismologically inferred phase speed. This effect has been discussed in previous CoMP studies. For example, \cite{2015NatCo...Morton} proposed that the flow speed can be estimated by separately measuring the phase speeds of outward- and inward-propagating waves. In CoMP and UCoMP observations, both outward and inward waves can in principle be identified through wave-tracking analyses because wave reflections are present \citep[e.g.,][]{2009ApJ...Tomczyk,2015NatCo...Morton,2024Sci...386...76Y}. Since the apparent phase speeds of the two wave components are modified in opposite ways by the background flow, the flow speed can be estimated from their difference. Therefore, in CoMP/UCoMP observations, the flow speed can in principle be constrained and can in principle be accounted for, at least in open-field regions, to obtain more accurate magnetic-field measurements.

In the present model, however, wave reflections are negligible or very weak, so outward-propagating waves overwhelmingly dominate. Under these conditions, it is not feasible to infer the flow speed from the wave-tracking procedure alone. To account for this effect in our comparison, we include the flow-speed term in the reference speed calculation, as in Eq. \ref{eq:vph_weight} and Figure \ref{fig:vph}. We also note that the $\epsilon$-weighted $B_\text{POS}$ shown in Figure \ref{fig:bfield}(B) does not account for the contribution from the background flow. This may partly explain why the measured magnetic field is slightly overestimated in the green-masked regions in Figure \ref{fig:bfield}(D) compared to the reference field in Figure \ref{fig:bfield}(B).

\section{Summary}

In this work, we investigate what physical quantities are actually measured by 2D coronal seismology. Using synthetic CoMP/UCoMP-like observables constructed from a global MHD model with a much longer LOS extent and more complex coronal structures than those used in previous forward-modeling studies, we investigated how LOS integration affects the physical interpretation of the quantities measured by 2D coronal seismology. By comparing the diagnosed density, wave propagation direction, phase speed, and magnetic-field strength with different model reference quantities, we evaluated what physical quantities these measurements represent under realistic optically thin LOS integration. The results show that the density diagnosed from the Fe \textsc{xiii} line ratio and the wave propagation direction derived from Doppler-velocity wave tracking are both well represented by LOS emissivity-weighted quantities. These two diagnostics remain robust even in the presence of substantial LOS complexity.

The interpretation of the phase speed and magnetic-field strength measurements is, however, more complicated. Unlike the density and wave propagation direction, the measured phase speed is generally not determined by emissivity weighting alone, because the wave-tracking procedure preferentially follows the LOS structures that produce the clearest propagating Doppler-velocity signatures. As a result, the LOS distribution of velocity amplitude can become an important additional factor in the effective LOS weighting. In the present model, an effective propagation speed weighted by $\epsilon^2A^2$ provides a significantly closer match to the measured phase speed than emissivity weighting alone, and the magnetic field inferred from Equation~\ref{eq:kinkeq} is correspondingly better approximated by combining the $\epsilon$-weighted density with the $\epsilon^2A^2$-weighted propagation speed. At the same time, our analysis also shows that simple emissivity weighting remains a reasonable approximation over a substantial part of the FOV, especially where the LOS does not intersect structures with strong variations in velocity amplitude. Previous studies have shown that the amplitudes of kink-like oscillations and waves in different coronal regions typically vary within about one order of magnitude. This suggests that, for many coronal regions, simple emissivity weighting is still likely to provide a useful first-order description for interpreting 2D coronal seismology measurements.

It is also important to place these results in the observational context. In the global MHD model used here, the LOS frequently crosses large-scale plasma-sheet and streamer-cusp structures, making the present setup a relatively extreme case for testing LOS effects. In actual CoMP/UCoMP observations, such large-scale structures are often located near or above the upper boundary of the typical coronal magnetic-field measurement FOV, which may reduce their impact on many practical measurements. However, their influence does not necessarily disappear, since LOS projection can extend the contribution of high-altitude structures to lower projected heights. In addition, smaller-scale plasma-sheet-like structures may still affect the interpretation of measured phase speeds and magnetic fields in localized regions. Therefore, while amplitude-dependent LOS effects may be reduced in typical CoMP/UCoMP applications compared with the configuration examined here, they can remain important in regions containing complex LOS structures. Overall, LOS emissivity weighting remains a useful first-order interpretation for much of the observed corona, but deviations associated with amplitude-dependent weighting should be considered in such structured regions.

More broadly, because 2D coronal seismology currently provides a 2D result formed through optically thin LOS integration, an important objective of this work is to identify a physically meaningful and reasonably accurate relationship between the measured magnetic-field-related quantities and the LOS distributions of relevant coronal properties, including emissivity, magnetic-field strength, velocity amplitude, and density. Establishing such a relationship is a key step toward future tomographic applications of 2D coronal seismology aimed at reconstructing the three-dimensional coronal magnetic-field distribution \citep[e.g., using the neural-field tomography framework as described in][]{2024ApJ...Jarolim}. At the same time, a clearer physical interpretation of the measurements will also allow 2D coronal seismology to provide stronger constraints on existing coronal magnetic-field models and to help refine the models.

\begin{acknowledgments}
This material is based upon work supported
by the NSF National Center for Atmospheric Research, which is a major facility
sponsored by the U.S. National Science Foundation under Cooperative Agreement No.
1852977. We would like to acknowledge high-performance computing support from the Derecho system (doi:10.5065/qx9a-pg09) provided by the NSF National Center for Atmospheric Research (NCAR), sponsored by the National Science Foundation. CHIANTI is a collaborative project involving George Mason University, the University of Michigan (USA), University of Cambridge (UK) and NASA Goddard Space Flight Center (USA). 

\end{acknowledgments}

\software{CHIANTI}

\appendix
\section{Justification for the phase-speed weighting}\label{app:vph_weight}

In the main text, we compare the measured quantities from 2D coronal seismology with several LOS-weighted reference quantities and find that different diagnostics are associated with different effective weighting factors. In particular, the phase-speed measurement can depend on the LOS distribution of both velocity amplitude and emissivity. In this appendix, we provide simple analytic arguments to justify the result.

To understand the effective LOS weighting in the phase-speed measurement, we consider two neighboring positions, $x_1$ and $x_2$, along the wave-tracking path. For the $i$th LOS element, we write the local Doppler-velocity perturbation at position $x$ as
\begin{equation}
    v_i(x,t)=A_i(x)\,\psi\!\left[t-T_i(x)\right],
\end{equation}
where $A_i(x)$ is the local LOS-projected velocity amplitude, $T_i(x)$ is the propagation delay of that LOS element, and $\psi$ is a common normalized waveform.

Under optically thin conditions, the observed Doppler signal at each position is the LOS superposition of all contributing elements weighted by their emissivities \citep[e.g.,][]{2026NatAs..10...42M}. To first order in the Doppler perturbation, the measured Doppler time series at the two neighboring positions can therefore be written as
\begin{equation}
    V_1(t)=\sum_{i=1}^{N}\epsilon_{1,i}A_{1,i}\,\psi(t-T_{1,i}),
\end{equation}
\begin{equation}
    V_2(t)=\sum_{i=1}^{N}\epsilon_{2,i}A_{2,i}\,\psi(t-T_{2,i}),
\end{equation}
where $\epsilon_{1,i}$ and $\epsilon_{2,i}$ are the emissivity contributions of the $i$th LOS element at $x_1$ and $x_2$, respectively, and similarly for $A_{1,i}$, $A_{2,i}$, $T_{1,i}$, and $T_{2,i}$.

The measured phase speed is inferred from the time lag between the Doppler signals at neighboring positions. For simplicity, we represent this procedure by the cross-correlation
\begin{equation}
    C_{12}(\tau)=\left< V_1(t)\,V_2(t+\tau)\right>.
\end{equation}
Substituting the two LOS-integrated signals gives
\begin{equation}
    C_{12}(\tau)=
    \sum_{i=1}^{N}\sum_{j=1}^{N}
    \epsilon_{1,i}\epsilon_{2,j}A_{1,i}A_{2,j}
    \left<
    \psi(t-T_{1,i})\,
    \psi(t+\tau-T_{2,j})
    \right>.
\end{equation}
here $i$ and $j$ correspond to two locations along the same LOS.

Because all LOS elements are assumed to share the same waveform $\psi$, the quantity in angle brackets can be written as the autocorrelation function of $\psi$ evaluated at the corresponding lag,
\begin{equation}
    R_\psi\!\left[\tau-(T_{2,j}-T_{1,i})\right]
    \equiv
    \left<
    \psi(t-T_{1,i})\,
    \psi(t+\tau-T_{2,j})
    \right>.
\end{equation}
Therefore,
\begin{equation}
    C_{12}(\tau)=
    \sum_{i=1}^{N}\sum_{j=1}^{N}
    \epsilon_{1,i}\epsilon_{2,j}A_{1,i}A_{2,j}
    R_\psi\!\left[\tau-(T_{2,j}-T_{1,i})\right].
\end{equation}

Equation above is still fully general under the simplified assumption of a common waveform. To proceed further, we make an additional physical assumption: signals contributed by different LOS elements are generally not phase-coherent over the time interval used in the cross-correlation, unless they arise from the same underlying propagating structure sampled at the two neighboring positions. In other words, for $i\neq j$, the corresponding contributions are expected to be reduced by loss of coherence, different phases, or different local wave evolution along distinct LOS structures. As a result, after temporal averaging, the off-diagonal terms make a much smaller contribution than the terms that correlate the same coherent structure between the two neighboring positions.

Under this assumption, the dominant contribution to the measured propagating branch is approximately given by the $i=j$ terms,
\begin{equation}
    C_{12}(\tau)\approx
    \sum_{i=1}^{N}
    \epsilon_{1,i}\epsilon_{2,i}A_{1,i}A_{2,i}
    R_\psi(\tau-\Delta T_i),
\end{equation}
where
\begin{equation}
    \Delta T_i \equiv T_{2,i}-T_{1,i}.
\end{equation}

This form shows that each coherent LOS element contributes a correlation peak centered near its own propagation delay $\Delta T_i$, with a strength proportional to
\begin{equation}
    \epsilon_{1,i}\epsilon_{2,i}A_{1,i}A_{2,i}.
\end{equation}
If the wave properties vary smoothly over the short spatial separation between the two neighboring positions, then it is reasonable to approximate
\begin{equation}
    \epsilon_{1,i}\approx \epsilon_{2,i}\approx \epsilon_i,
    \qquad
    A_{1,i}\approx A_{2,i}\approx A_i,
\end{equation}
which reduces the effective weight of the $i$th LOS element to
\begin{equation}
    \epsilon_i^2 A_i^2.
\end{equation}

In the special case where the dominant $\Delta T_i$ values are close to one another, the measured lag is then approximately
\begin{equation}
    \tau_{\rm meas}\approx
    \frac{\sum_{i\in {\rm dom}}\epsilon_i^2A_i^2\,\Delta T_i}
    {\sum_{i\in {\rm dom}}\epsilon_i^2A_i^2}.
\end{equation}
When different groups of LOS elements contribute at distinct delays, the selected branch is expected to be controlled by the group with the largest combined contribution, approximately measured by $\sum_{i\in g}\epsilon_i^2A_i^2$.

This simplified argument is not intended to establish a unique universal weighting law for all wave-tracking procedures. Rather, it shows why the phase-speed measurement, unlike the density diagnostic, naturally depends on both emissivity and velocity amplitude at two neighboring positions, making a weighting of the form $\epsilon^2A^2$ physically well motivated. This is also consistent with the forward-modeling results in the main text, where the $\epsilon^2A^2$ weighting provides the closest empirical match to the measured phase speed.

\section{An Idealized Toy Model for Assessing Amplitude-weighting Effects}\label{ap:toymodel}

To isolate the role of amplitude weighting in the phase-speed measurement, we construct an idealized toy LOS model and apply the same wave-tracking procedure used in the main text. The purpose of this model is not to reproduce the full complexity of the global MHD simulation, but to provide a controlled setup in which the LOS distributions of phase speed, emissivity, and velocity amplitude can be prescribed independently.

\begin{figure}
    \centering
    \includegraphics[width=0.9\linewidth]{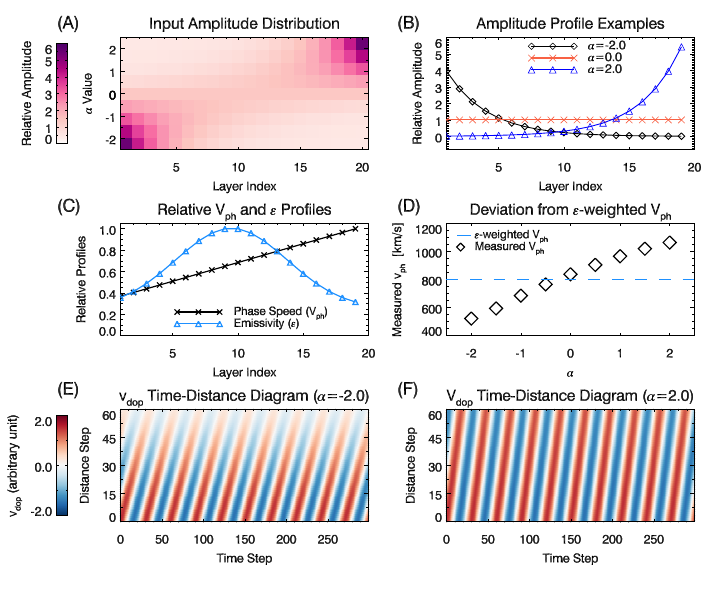}
    \caption{Setup and analysis results of the toy model. (A) Distribution of the prescribed input amplitude. The x-axis shows the layer index, the y-axis is the $\alpha$ value, and the colors represent the relative amplitude. (B) Three examples of the amplitude profile across the LOS layers. The black diamonds, red crosses, and blue triangles correspond to the cases with $\alpha=-2$, 0, and 2, respectively. (C) Profiles of the relative phase speed (black crosses) and emissivity (blue triangles) across the LOS layers. (D) Measured phase speeds for all nine cases (black diamonds), together with the LOS emissivity-weighted phase speed (blue dashed line). (E) Example of the LOS-integrated Doppler-velocity time--distance diagram for the case with $\alpha=-2$. (F) Same as panel (E), but for the case with $\alpha=2$.}
    \label{fig:toymodel}
\end{figure}

We represent the LOS by $N_{\rm LOS}=20$ discrete layers, labeled by index $i$. A normalized LOS coordinate is defined as
\begin{equation}
s_i = -1 + \frac{2(i-1)}{N_{\rm LOS}-1},
\qquad i=1,\dots,N_{\rm LOS},
\end{equation}
so that $s_i=-1$ and $+1$ correspond to the two ends of the LOS. The phase speed is assumed to vary monotonically across the LOS between 400 km s$^{-1}$ and 1200 km s$^{-1}$,
\begin{equation}
v_{{\rm ph},i}
=
400
+
\frac{i-1}{N_{\rm LOS}-1}(1200-400)
\ {\rm km\ s^{-1}},
\end{equation}
such that one side of the LOS is dominated by slower waves and the other by faster waves. The black crosses in Figure \ref{fig:toymodel}(C) show the normalized phase-speed profile across the LOS layers.

The emissivity profile is fixed in all cases and is chosen to peak near the middle LOS layers:
\begin{equation}
\epsilon_i \propto 0.25 + 0.75\exp\left[-\frac12\left(\frac{s_i}{0.45}\right)^2\right].
\end{equation}
After construction, the emissivity profile is normalized to have unit mean across the LOS. The resulting profile is shown by the blue triangles in Figure \ref{fig:toymodel}(C).

All layers are assumed to oscillate with the same period,
\begin{equation}
P=150\ {\rm s},
\qquad
\omega=\frac{2\pi}{P},
\end{equation}
and all layers are initialized with the same phase, so that the effect of phase offsets between LOS components is intentionally suppressed.

For the $i$th LOS layer, we prescribe the local LOS-projected velocity perturbation as
\begin{equation}
u_i(x,t)=A_i
\cos\left(\omega t-\frac{\omega x}{v_{{\rm ph},i}}+\phi_{0}\right),
\end{equation}
where $A_i$ is the LOS-projected velocity amplitude of the $i$th layer and $\phi_{0}=0$ in the present setup. Under optically thin conditions, the observed Doppler contribution from that layer is weighted by its emissivity, so that
\begin{equation}
V_i(x,t)=\epsilon_i\,u_i(x,t).
\end{equation}
The total synthetic Doppler signal is then obtained by summing all LOS contributions,
\begin{equation}
V(x,t)=\sum_{i=1}^{N_{\rm LOS}} V_i(x,t).
\end{equation}
Because the model is intended only to test LOS weighting effects, the wave properties are taken to be uniform in the transverse direction $y$.

The synthetic data cube contains 201 spatial pixels along $x$, 21 pixels along $y$, and 300 time steps. We adopt a spatial sampling of $\Delta x=\Delta y=1.5$ Mm and a temporal cadence of $\Delta t=5$ s. The phase speed is measured from the central $y$ row using the same wave-tracking procedure as in the main text, applied to the synthesized time--distance diagram (e.g., Figure \ref{fig:toymodel}(E) and (F)).

To test the impact of velocity amplitude, we prescribe a monotonic amplitude variation across the LOS:
\begin{equation}
A_i \propto e^{1.5\,\alpha\, s_i},
\end{equation}
where $\alpha$ controls the degree and direction of the asymmetry. We consider nine cases with $\alpha=-2,-1.5,-1,-0.5,0,0.5,1,1.5,2$. In all cases, $A_i$ is normalized to have unit mean across the LOS, so that the comparison isolates the effect of the amplitude distribution rather than the absolute amplitude scale. The resulting amplitude distributions for all nine cases are shown in Figure \ref{fig:toymodel}(A). Figure \ref{fig:toymodel}(B) further presents three representative examples corresponding to $\alpha=-2$, 0, and 2.

If amplitude weighting were unimportant, the measured phase speeds from the nine cases would be expected to remain similar despite the different LOS amplitude profiles. Instead, the time--distance diagrams in Figure \ref{fig:toymodel}(E) and (F) show substantially different propagation slopes. Moreover, Figure \ref{fig:toymodel}(D) compares the measured phase speed from each case (black diamonds) with the emissivity-weighted reference speed (blue dashed line). As $\alpha$ increases from $-2$ to 2, the measured phase speed shifts systematically from values below the $\epsilon$-weighted reference to values above it. This provides clear evidence that the LOS distribution of velocity amplitude can affect the wave-tracking procedure and thereby modify the measured phase speed, and consequently the inferred magnetic-field strength.

This toy model therefore provides a controlled demonstration of how asymmetric amplitude distributions along the LOS can shift the measured phase speed away from a simple emissivity-weighted average.

\section{Additional Examples under Different Viewing Angles}\label{ap:add}

In the main text, we present results for the case with a $B$-angle=$10^\circ$ rotation of the MHD model. To further examine the effect of viewing angle and LOS complexity, we also analyze two additional cases with $B$-angle=$0^\circ$ and $20^\circ$. Both cases show the same overall behavior as the main-text example.

\begin{figure}
    \centering   \includegraphics[width=0.75\linewidth]{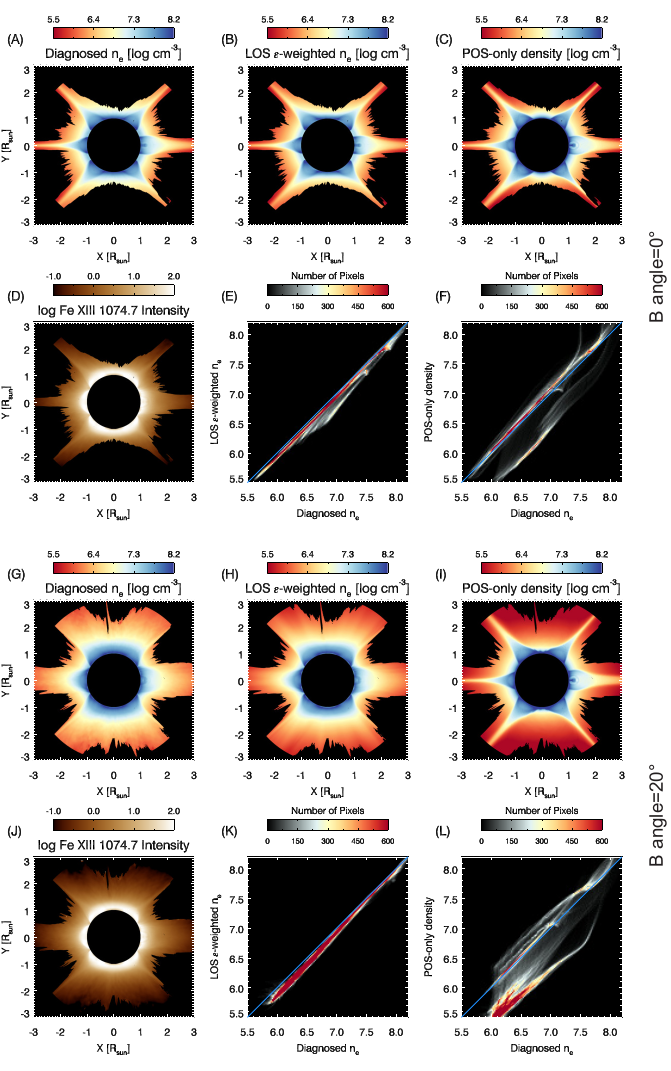}
    \caption{Same as Figure \ref{fig:density}, but for two additional viewing-angle cases with model rotations of $B$-angle=$0^\circ$ (panels (A)--(F)) and $B$-angle=$20^\circ$ (panels (G)--(L)).}
    \label{fig:dens_append}
\end{figure}

Figure \ref{fig:dens_append} is similar to Figure \ref{fig:density} and shows that the diagnosed density can still be well approximated by the LOS $\epsilon$-weighted average.

\begin{figure}
    \centering    \includegraphics[width=0.9\linewidth]{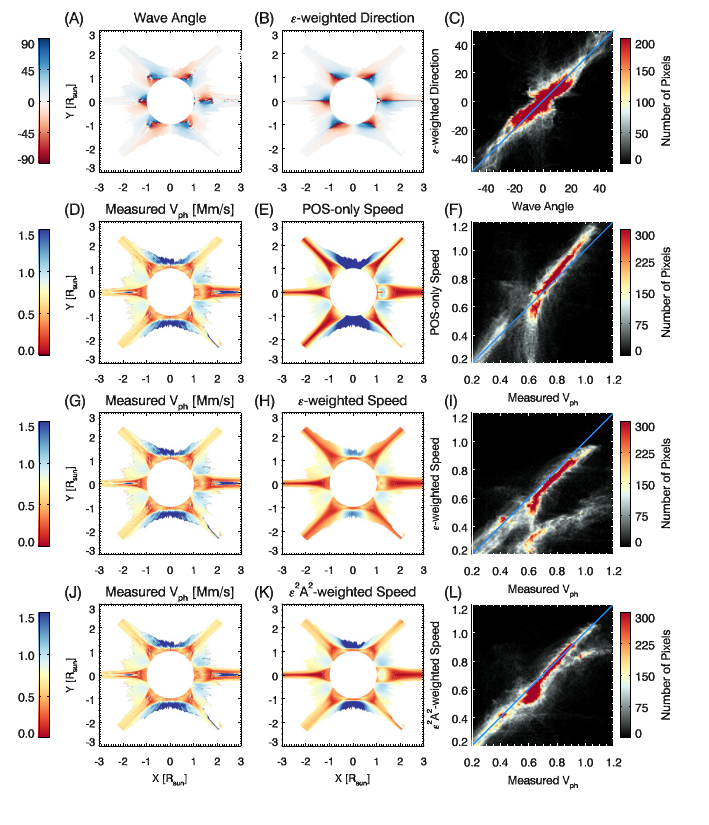}
    \caption{Same as Figure \ref{fig:vph}, but for the additional viewing-angle case with a model rotation of $B$-angle=$0^\circ$.}
    \label{fig:vph_theta0}
\end{figure}

\begin{figure}[htbp]
    \centering    \includegraphics[width=0.9\linewidth]{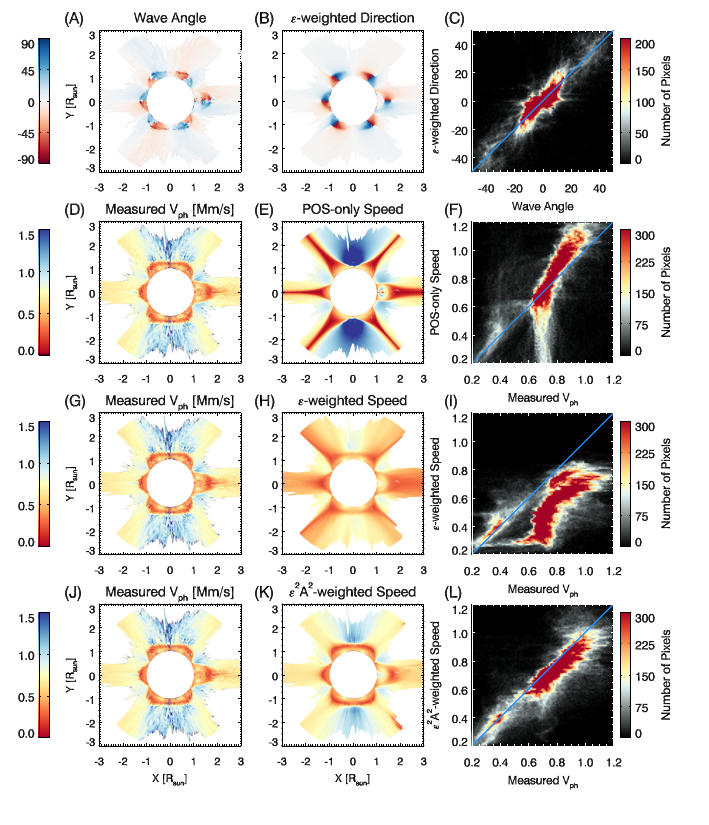}
    \caption{Same as Figure \ref{fig:vph}, but for the additional viewing-angle case with a model rotation of $B$-angle=$20^\circ$.}
    \label{fig:vph_theta20}
\end{figure}

Figures \ref{fig:vph_theta0} and \ref{fig:vph_theta20} show the phase-speed results for the cases with $B$-angle=$0^\circ$ and $20^\circ$, respectively. Both figures indicate that the measured phase speed depends not only on emissivity weighting but also on the LOS distribution of velocity amplitude. Compared with the $B$-angle=$20^\circ$ case, the $B$-angle=$0^\circ$ case shows better agreement between the measured phase speed and the emissivity-only weighted speed. This is because the LOS structure is more symmetric for $B$-angle=$0^\circ$, so only a relatively small fraction of the FOV contains LOS that crosses plasma-sheet regions.

\begin{figure}
    \centering  \includegraphics[width=0.99\linewidth]{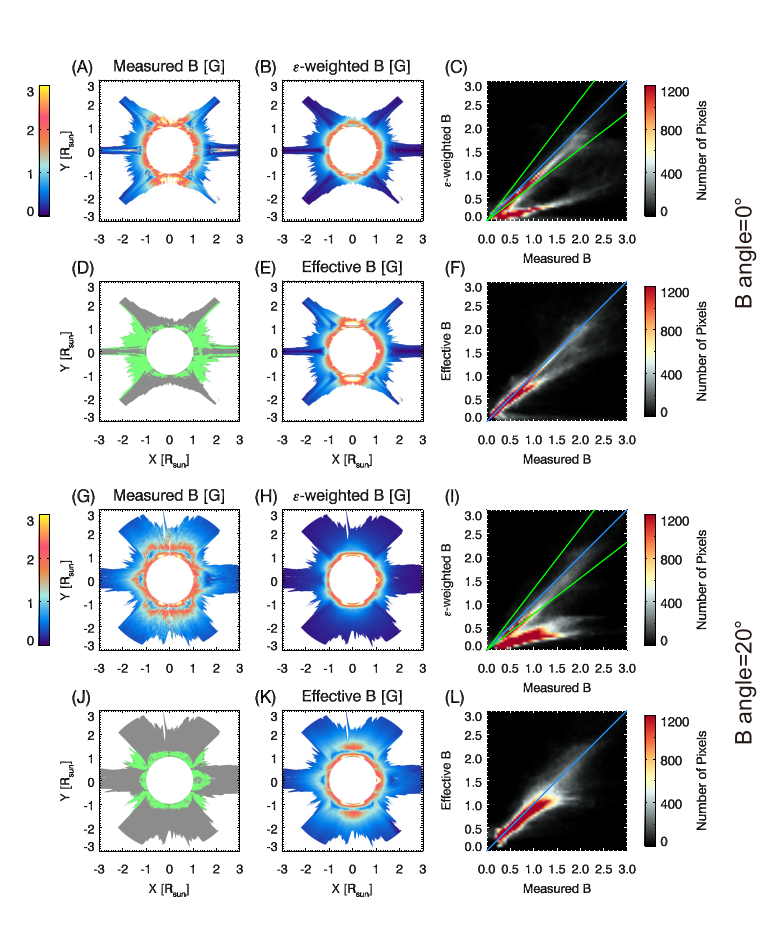}
    \caption{Same as Figure \ref{fig:bfield}, but for the additional viewing-angle cases with model rotations of $B$-angle=$0^\circ$ (panels (A) to (F)) and $B$-angle=$20^\circ$ (panels (G) to (L)).}
    \label{fig:bmap_append}
\end{figure}

Figure \ref{fig:bmap_append} shows the magnetic-field comparison for both additional viewing-angle cases, similar to Figure \ref{fig:bfield}. In both cases, the measured magnetic field is better represented by the effective magnetic-field strength defined in Equation \ref{eq:beff}. The green contours in panels (A) and (G) mark the regions where simple emissivity weighting still provides a reasonable approximation. As discussed in the main text, low-height and open-field regions without plasma-sheet structures along the LOS can in general still be approximated by simple emissivity weighting.

\section{Additional Examples of the Impact of Velocity Amplitude on LOS Weighting}\label{sec:app_weight}

\begin{figure}
    \centering
    \includegraphics[width=0.9\linewidth]{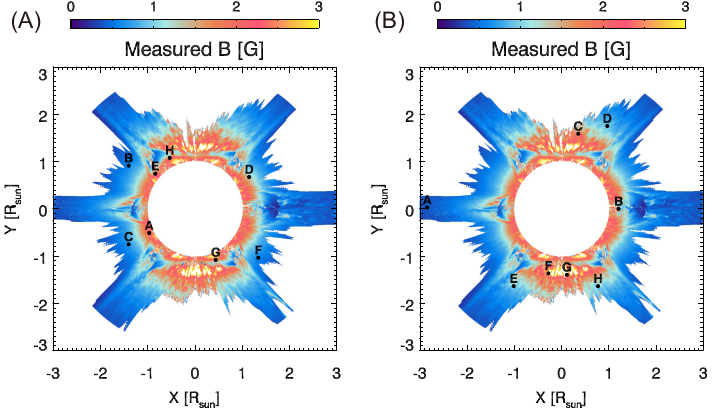}
    \caption{Measured magnetic-field maps for the case with a model rotation of $B$-angle=$10^\circ$. (A) The dots with letters mark locations where emissivity weighting remains a good approximation to the measured magnetic field. (B) The dots with letters correspond to locations where emissivity-weighting fails and velocity amplitude becomes important in the LOS weighting.}
    \label{fig:bmap_weight_app}
\end{figure}

\begin{figure}
    \centering
    \includegraphics[width=0.99\linewidth]{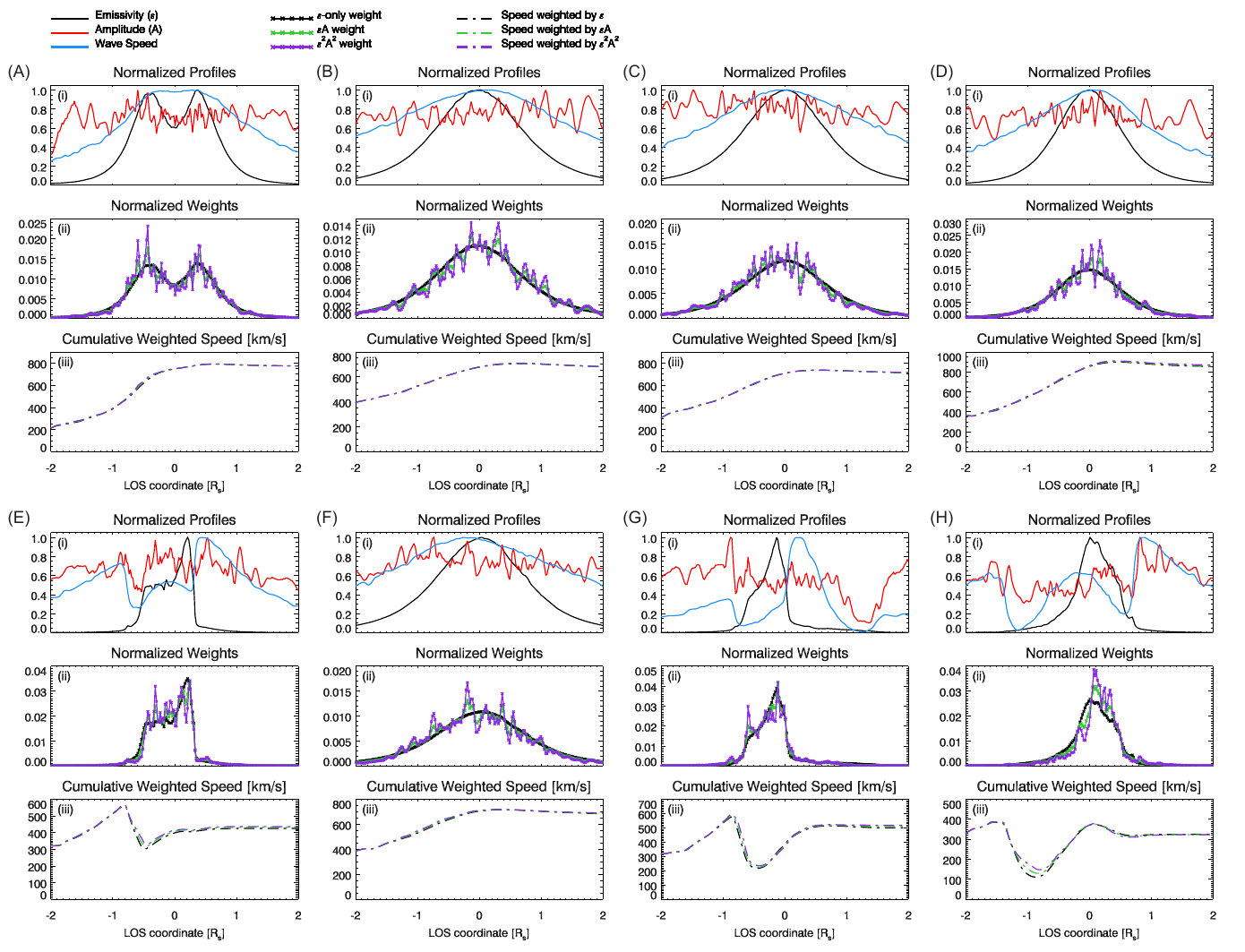}
    \caption{Similar to Figure \ref{fig:weights}, but for locations marked by letters in Figure \ref{fig:bmap_weight_app}(A), where emissivity-weighting is still a good approximation to the measurements.}
    \label{fig:weight_profile_good}
\end{figure}

\begin{figure}
    \centering
    \includegraphics[width=0.99\linewidth]{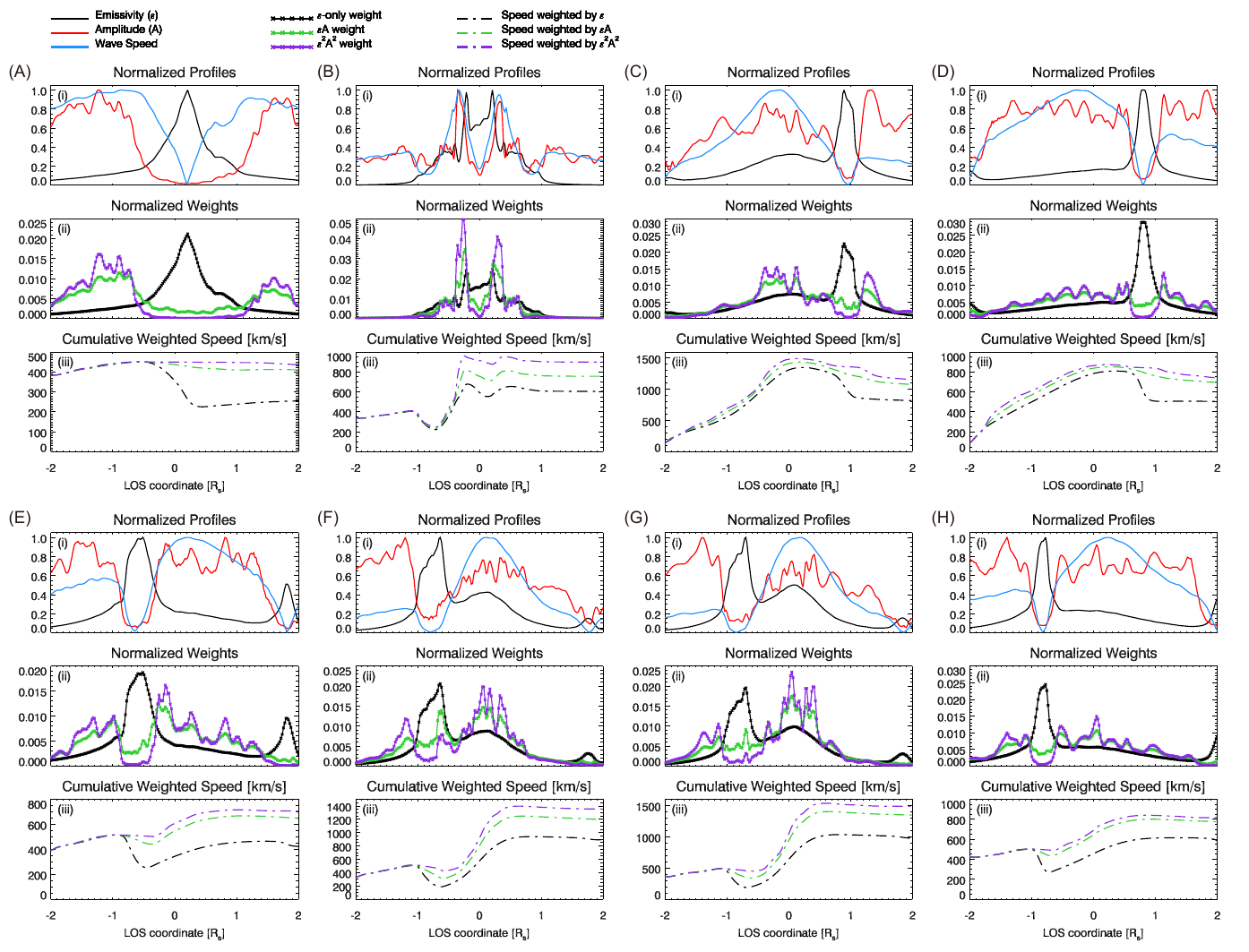}
    \caption{Similar to Figure \ref{fig:weight_profile_good}, but for locations marked by letters in Figure \ref{fig:bmap_weight_app}(B), where velocity amplitude becomes important in LOS weighting.}
    \label{fig:weight_profile_bad}
\end{figure}

In this appendix, we show additional examples of LOS weighting in different regions. Figure \ref{fig:bmap_weight_app} shows the measured magnetic field for the case with a model rotation of $B$-angle=$10^\circ$, the same viewing angle as in the main text. The letters in Figure~\ref{fig:bmap_weight_app}(A) mark locations where emissivity weighting provides a good approximation to the measured magnetic field, and the corresponding LOS distributions of velocity amplitude and their effects on LOS weighting are shown in Figure \ref{fig:weight_profile_good}. Similarly, Figure \ref{fig:bmap_weight_app}(B) and Figure \ref{fig:weight_profile_bad} show locations where velocity amplitude becomes an important additional weighting factor. All cases show that, in regions where simple emissivity weighting fails, the corresponding LOS tend to cross plasma-sheet regions where emissivity is anti-correlated with velocity amplitude and phase speed. 

There is one region of particular interest. In the low-lying western streamer, where the LOS intersects the prominence-cavity system, a relatively large discrepancy is found between the measured and emissivity-weighted quantities (marked by A in Figure \ref{fig:bmap_weight_app}(B)). The LOS profiles in Figure \ref{fig:weight_profile_bad}(B) show that the emissivity, wave amplitude, and local wave speed are strongly structured along the LOS, with their peaks occurring at different locations. As a result, emissivity-only weighting gives substantial weight to bright plasma with relatively low wave speeds, whereas the inclusion of wave-amplitude weighting increasingly emphasizes neighboring components with higher wave speeds. This suggests that the discrepancy is primarily associated with the complex LOS structure of the prominence-cavity environment.

\clearpage
\bibliography{sample701}{}

@ARTICLE{2008SoPh..247..411T,
       author = {{Tomczyk}, S. and {Card}, G.~L. and {Darnell}, T. and {Elmore}, D.~F. and {Lull}, R. and {Nelson}, P.~G. and {Streander}, K.~V. and {Burkepile}, J. and {Casini}, R. and {Judge}, P.~G.},
        title = "{An Instrument to Measure Coronal Emission Line Polarization}",
      journal = {\solphys},
         year = 2008,
        month = feb,
       volume = {247},
       number = {2},
        pages = {411-428},
          doi = {10.1007/s11207-007-9103-6},
       adsurl = {https://ui.adsabs.harvard.edu/abs/2008SoPh..247..411T}
}

@ARTICLE{2026ApJS..283....3Y,
       author = {{Yang}, Zihao and {Rempel}, Matthias and {Gibson}, Sarah and {de Toma}, Giuliana},
        title = "{Measuring the Coronal Magnetic Field with 2D Coronal Seismology: A Forward-modeling Validation}",
      journal = {\apjs},
         year = 2026,
        month = mar,
       volume = {283},
       number = {1},
          eid = {3},
        pages = {3},
          doi = {10.3847/1538-4365/ae39c5},
archivePrefix = {arXiv},
       eprint = {2601.10637},
 primaryClass = {astro-ph.SR},
       adsurl = {https://ui.adsabs.harvard.edu/abs/2026ApJS..283....3Y}
}

@ARTICLE{1997A&AS..125..149D,
       author = {{Dere}, K.~P. and {Landi}, E. and {Mason}, H.~E. and {Monsignori Fossi}, B.~C. and {Young}, P.~R.},
        title = "{CHIANTI - an atomic database for emission lines}",
      journal = {\aaps},
         year = 1997,
        month = oct,
       volume = {125},
        pages = {149-173},
          doi = {10.1051/aas:1997368},
       adsurl = {https://ui.adsabs.harvard.edu/abs/1997A&AS..125..149D}
}

@ARTICLE{2024ApJ...974...71D,
       author = {{Dufresne}, R.~P. and {Del Zanna}, G. and {Young}, P.~R. and {Dere}, K.~P. and {Deliporanidou}, E. and {Barnes}, W.~T. and {Landi}, E.},
        title = "{CHIANTI{\textemdash}An Atomic Database for Emission Lines{\textemdash}Paper. XVIII. Version 11, Advanced Ionization Equilibrium Models: Density and Charge Transfer Effects}",
      journal = {\apj},
         year = 2024,
        month = oct,
       volume = {974},
       number = {1},
          eid = {71},
        pages = {71},
          doi = {10.3847/1538-4357/ad6765},
archivePrefix = {arXiv},
       eprint = {2403.16922},
 primaryClass = {astro-ph.SR},
       adsurl = {https://ui.adsabs.harvard.edu/abs/2024ApJ...974...71D}
}

@ARTICLE{2024ApJ...966..122Z,
       author = {{Zhu}, Yingjie and {Habbal}, Shadia R. and {Ding}, Adalbert and {Yamashiro}, Bryan and {Landi}, Enrico and {Boe}, Benjamin and {Constantinou}, Sage and {Nassir}, Michael},
        title = "{Spectroscopic Observations of the Solar Corona during the 2017 August 21 Total Solar Eclipse: Comparison of Spectral Line Widths and Doppler Shifts between Open and Closed Magnetic Structures}",
      journal = {\apj},
         year = 2024,
        month = may,
       volume = {966},
       number = {1},
          eid = {122},
        pages = {122},
          doi = {10.3847/1538-4357/ad3424},
archivePrefix = {arXiv},
       eprint = {2403.10363},
 primaryClass = {astro-ph.SR},
       adsurl = {https://ui.adsabs.harvard.edu/abs/2024ApJ...966..122Z}
}

@ARTICLE{2009A&A...501L..15B,
       author = {{Banerjee}, D. and {P{\'e}rez-Su{\'a}rez}, D. and {Doyle}, J.~G.},
        title = "{Signatures of Alfv{\'e}n waves in the polar coronal holes as seen by EIS/Hinode}",
      journal = {\aap},
         year = 2009,
        month = jul,
       volume = {501},
       number = {3},
        pages = {L15-L18},
          doi = {10.1051/0004-6361/200912242},
archivePrefix = {arXiv},
       eprint = {0906.4600},
 primaryClass = {astro-ph.SR},
       adsurl = {https://ui.adsabs.harvard.edu/abs/2009A&A...501L..15B}
}

@ARTICLE{2011ApJ...727L..13B,
       author = {{Brooks}, David H. and {Warren}, Harry P.},
        title = "{Establishing a Connection Between Active Region Outflows and the Solar Wind: Abundance Measurements with EIS/Hinode}",
      journal = {\apjl},
         year = 2011,
        month = jan,
       volume = {727},
       number = {1},
          eid = {L13},
        pages = {L13},
          doi = {10.1088/2041-8205/727/1/L13},
archivePrefix = {arXiv},
       eprint = {1009.4291},
 primaryClass = {astro-ph.SR},
       adsurl = {https://ui.adsabs.harvard.edu/abs/2011ApJ...727L..13B}
}

@INPROCEEDINGS{2021AGUFMSH15G2089T,
       author = {{Tomczyk}, Steven and {Landi}, Enrico and {Berkey}, Ben and {Burkepile}, Joan and {Cotter}, Marc and {Gallaher}, Dennis and {Galloy}, Michael D. and {Graves}, Rob and {Oakley}, Philip and {Perez-Gonzalez}, Lisa and {Sewell}, Scott and {de Toma}, Giuliana and {Zmarzly}, Patrick},
        title = "{First Images from the Upgraded Coronal Multi-channel Polarimeter (UCoMP)}",
    booktitle = {AGU Fall Meeting Abstracts},
         year = 2021,
       series = {AGU Fall Meeting Abstracts},
       volume = {2021},
        month = dec,
        pages = {2089},
       adsurl = {https://ui.adsabs.harvard.edu/abs/2021AGUFMSH15G2089T}
}

@ARTICLE{2024Sci...386...76Y,
       author = {{Yang}, Zihao and {Tian}, Hui and {Tomczyk}, Steven and {Liu}, Xianyu and {Gibson}, Sarah and {Morton}, Richard J. and {Downs}, Cooper},
        title = "{Observing the evolution of the Sun's global coronal magnetic field over 8 months}",
      journal = {Science},
         year = 2024,
        month = oct,
       volume = {386},
       number = {6717},
        pages = {76-82},
          doi = {10.1126/science.ado2993},
archivePrefix = {arXiv},
       eprint = {2410.16555},
 primaryClass = {astro-ph.SR},
       adsurl = {https://ui.adsabs.harvard.edu/abs/2024Sci...386...76Y}
}

@ARTICLE{2013SoPh..288..637T,
       author = {{Tian}, H. and {Tomczyk}, S. and {McIntosh}, S.~W. and {Bethge}, C. and {de Toma}, G. and {Gibson}, S.},
        title = "{Observations of Coronal Mass Ejections with the Coronal Multichannel Polarimeter}",
      journal = {\solphys},
         year = 2013,
        month = dec,
       volume = {288},
       number = {2},
        pages = {637-650},
          doi = {10.1007/s11207-013-0317-5},
archivePrefix = {arXiv},
       eprint = {1303.4647},
 primaryClass = {astro-ph.SR},
       adsurl = {https://ui.adsabs.harvard.edu/abs/2013SoPh..288..637T}
}

@ARTICLE{2024ApJ...963...69B,
       author = {{Baweja}, Upasna and {Pant}, Vaibhav and {Arregui}, I{\~n}igo},
        title = "{Global Coronal Magnetic Field Estimation Using Bayesian Inference}",
      journal = {\apj},
         year = 2024,
        month = mar,
       volume = {963},
       number = {1},
          eid = {69},
        pages = {69},
          doi = {10.3847/1538-4357/ad1b57},
archivePrefix = {arXiv},
       eprint = {2401.05022},
 primaryClass = {astro-ph.SR},
       adsurl = {https://ui.adsabs.harvard.edu/abs/2024ApJ...963...69B}
}

@ARTICLE{2020Sci...369..694Y,
       author = {{Yang}, Zihao and {Bethge}, Christian and {Tian}, Hui and {Tomczyk}, Steven and {Morton}, Richard and {Del Zanna}, Giulio and {McIntosh}, Scott W. and {Karak}, Bidya Binay and {Gibson}, Sarah and {Samanta}, Tanmoy and {He}, Jiansen and {Chen}, Yajie and {Wang}, Linghua},
        title = "{Global maps of the magnetic field in the solar corona}",
      journal = {Science},
         year = 2020,
        month = aug,
       volume = {369},
       number = {6504},
        pages = {694-697},
          doi = {10.1126/science.abb4462},
archivePrefix = {arXiv},
       eprint = {2008.03136},
 primaryClass = {astro-ph.SR},
       adsurl = {https://ui.adsabs.harvard.edu/abs/2020Sci...369..694Y}
}

@ARTICLE{2020ScChE..63.2357Y,
       author = {{Yang}, ZiHao and {Tian}, Hui and {Tomczyk}, Steven and {Morton}, Richard and {Bai}, XianYong and {Samanta}, Tanmoy and {Chen}, YaJie},
        title = "{Mapping the magnetic field in the solar corona through magnetoseismology}",
      journal = {Science in China E: Technological Sciences},
         year = 2020,
        month = nov,
       volume = {63},
       number = {11},
        pages = {2357-2368},
          doi = {10.1007/s11431-020-1706-9},
archivePrefix = {arXiv},
       eprint = {2008.03146},
 primaryClass = {astro-ph.SR},
       adsurl = {https://ui.adsabs.harvard.edu/abs/2020ScChE..63.2357Y}
}

@ARTICLE{2021ApJ...Dudik,
       author = {{Dud{\'\i}k}, Jaroslav and {Del Zanna}, Giulio and {Ryb{\'a}k}, J{\'a}n and {L{\"o}rin{\v{c}}{\'\i}k}, Juraj and {Dzif{\v{c}}{\'a}kov{\'a}}, Elena and {Mason}, Helen E. and {Tomczyk}, Steven and {Galloy}, Michael},
        title = "{Electron Densities in the Solar Corona Measured Simultaneously in the Extreme Ultraviolet and Infrared}",
      journal = {\apj},
         year = 2021,
        month = jan,
       volume = {906},
       number = {2},
          eid = {118},
        pages = {118},
          doi = {10.3847/1538-4357/abcd91},
archivePrefix = {arXiv},
       eprint = {2011.09175},
 primaryClass = {astro-ph.SR},
       adsurl = {https://ui.adsabs.harvard.edu/abs/2021ApJ...906..118D}
}

@ARTICLE{2026ApJ...FeXIIINova,
       author = {{Banerjee}, D.~P.~K. and {Woodward}, C.~E. and {Evans}, A. and {Geballe}, T.~R. and {Joshi}, V. and {Starrfield}, S.},
        title = "{The [Fe XIII] Infrared 10747 and 10798 {\r{A}} Lines in Novae}",
      journal = {\apj},
         year = 2026,
        month = feb,
       volume = {998},
       number = {2},
          eid = {279},
        pages = {279},
          doi = {10.3847/1538-4357/ae3bc8},
archivePrefix = {arXiv},
       eprint = {2601.14218},
 primaryClass = {astro-ph.SR},
       adsurl = {https://ui.adsabs.harvard.edu/abs/2026ApJ...998..279B}
}

@ARTICLE{2020ApJ...Gilly,
       author = {{Gilly}, Chris R. and {Cranmer}, Steven R.},
        title = "{The Effect of Solar Wind Expansion and Nonequilibrium Ionization on the Broadening of Coronal Emission Lines}",
      journal = {\apj},
         year = 2020,
        month = oct,
       volume = {901},
       number = {2},
          eid = {150},
        pages = {150},
          doi = {10.3847/1538-4357/abb1ad},
archivePrefix = {arXiv},
       eprint = {2008.09580},
 primaryClass = {astro-ph.SR},
       adsurl = {https://ui.adsabs.harvard.edu/abs/2020ApJ...901..150G}
}

@ARTICLE{2024ApJ...faraday,
       author = {{Barnier}, Samuel and {Done}, Chris},
        title = "{Making the Invisible Visible: Magnetic Fields in Accretion Flows Revealed by X-Ray Polarization}",
      journal = {\apj},
         year = 2024,
        month = dec,
       volume = {977},
       number = {2},
          eid = {201},
        pages = {201},
          doi = {10.3847/1538-4357/ad9277},
archivePrefix = {arXiv},
       eprint = {2404.12815},
 primaryClass = {astro-ph.HE},
       adsurl = {https://ui.adsabs.harvard.edu/abs/2024ApJ...977..201B}
}

@ARTICLE{2021ApJ...yajie,
       author = {{Chen}, Yajie and {Li}, Wenxian and {Tian}, Hui and {Chen}, Feng and {Bai}, Xianyong and {Yang}, Yang and {Yang}, Zihao and {Liu}, Xianyu and {Deng}, Yuanyong},
        title = "{Forward Modeling of Solar Coronal Magnetic-field Measurements Based on a Magnetic-field-induced Transition in Fe X}",
      journal = {\apj},
         year = 2021,
        month = oct,
       volume = {920},
       number = {2},
          eid = {116},
        pages = {116},
          doi = {10.3847/1538-4357/ac1792},
archivePrefix = {arXiv},
       eprint = {2107.11783},
 primaryClass = {astro-ph.SR},
       adsurl = {https://ui.adsabs.harvard.edu/abs/2021ApJ...920..116C}
}

@ARTICLE{2015NatCo...Morton,
       author = {{Morton}, R.~J. and {Tomczyk}, S. and {Pinto}, R.},
        title = "{Investigating Alfv{\'e}nic wave propagation in coronal open-field regions}",
      journal = {Nature Communications},
         year = 2015,
        month = jul,
       volume = {6},
          eid = {7813},
        pages = {7813},
          doi = {10.1038/ncomms8813},
       adsurl = {https://ui.adsabs.harvard.edu/abs/2015NatCo...6.7813M}
}

@ARTICLE{2019NatAs...Morton,
       author = {{Morton}, R.~J. and {Weberg}, M.~J. and {McLaughlin}, J.~A.},
        title = "{A basal contribution from p-modes to the Alfv{\'e}nic wave flux in the Sun's corona}",
      journal = {Nature Astronomy},
         year = 2019,
        month = jan,
       volume = {3},
        pages = {223},
          doi = {10.1038/s41550-018-0668-9},
archivePrefix = {arXiv},
       eprint = {1902.03811},
 primaryClass = {astro-ph.SR},
       adsurl = {https://ui.adsabs.harvard.edu/abs/2019NatAs...3..223M}
}

@ARTICLE{2017A&A...Long,
       author = {{Long}, D.~M. and {Valori}, G. and {P{\'e}rez-Su{\'a}rez}, D. and {Morton}, R.~J. and {V{\'a}squez}, A.~M.},
        title = "{Measuring the magnetic field of a trans-equatorial loop system using coronal seismology}",
      journal = {\aap},
         year = 2017,
        month = jul,
       volume = {603},
          eid = {A101},
        pages = {A101},
          doi = {10.1051/0004-6361/201730413},
archivePrefix = {arXiv},
       eprint = {1703.10020},
 primaryClass = {astro-ph.SR},
       adsurl = {https://ui.adsabs.harvard.edu/abs/2017A&A...603A.101L}
}

@ARTICLE{2009ApJ...Tomczyk,
       author = {{Tomczyk}, Steven and {McIntosh}, Scott W.},
        title = "{Time-Distance Seismology of the Solar Corona with CoMP}",
      journal = {\apj},
         year = 2009,
        month = jun,
       volume = {697},
       number = {2},
        pages = {1384-1391},
          doi = {10.1088/0004-637X/697/2/1384},
archivePrefix = {arXiv},
       eprint = {0903.2002},
 primaryClass = {astro-ph.SR},
       adsurl = {https://ui.adsabs.harvard.edu/abs/2009ApJ...697.1384T}
}

@ARTICLE{2008SoPh..McIntosh,
       author = {{McIntosh}, Scott W. and {De Pontieu}, Bart and {Tomczyk}, Steven},
        title = "{A Coherence-Based Approach for Tracking Waves in the Solar Corona}",
      journal = {\solphys},
         year = 2008,
        month = nov,
       volume = {252},
       number = {2},
        pages = {321-348},
          doi = {10.1007/s11207-008-9257-x},
archivePrefix = {arXiv},
       eprint = {0808.2978},
 primaryClass = {astro-ph},
       adsurl = {https://ui.adsabs.harvard.edu/abs/2008SoPh..252..321M}
}

@ARTICLE{2025RAA....Gao,
       author = {{Gao}, Yuhang and {Tian}, Hui and {Van Doorsselaere}, Tom and {Yang}, Zihao and {Guo}, Mingzhe and {Karampelas}, Konstantinos},
        title = "{Measurements of the Solar Coronal Magnetic Field Based on Coronal Seismology with Propagating Alfv{\'e}nic Waves: Forward Modeling}",
      journal = {Research in Astronomy and Astrophysics},
         year = 2025,
        month = jan,
       volume = {25},
       number = {1},
          eid = {015010},
        pages = {015010},
          doi = {10.1088/1674-4527/ad981c},
archivePrefix = {arXiv},
       eprint = {2411.08310},
 primaryClass = {astro-ph.SR},
       adsurl = {https://ui.adsabs.harvard.edu/abs/2025RAA....25a5010G}
}

@ARTICLE{2012ApJ...Wang,
       author = {{Wang}, Tongjiang and {Ofman}, Leon and {Davila}, Joseph M. and {Su}, Yang},
        title = "{Growing Transverse Oscillations of a Multistranded Loop Observed by SDO/AIA}",
      journal = {\apjl},
         year = 2012,
        month = jun,
       volume = {751},
       number = {2},
          eid = {L27},
        pages = {L27},
          doi = {10.1088/2041-8205/751/2/L27},
archivePrefix = {arXiv},
       eprint = {1204.1376},
 primaryClass = {astro-ph.SR},
       adsurl = {https://ui.adsabs.harvard.edu/abs/2012ApJ...751L..27W}
}

@ARTICLE{2012ApJ...tian,
       author = {{Tian}, Hui and {McIntosh}, Scott W. and {Wang}, Tongjiang and {Ofman}, Leon and {De Pontieu}, Bart and {Innes}, Davina E. and {Peter}, Hardi},
        title = "{Persistent Doppler Shift Oscillations Observed with Hinode/EIS in the Solar Corona: Spectroscopic Signatures of Alfv{\'e}nic Waves and Recurring Upflows}",
      journal = {\apj},
         year = 2012,
        month = nov,
       volume = {759},
       number = {2},
          eid = {144},
        pages = {144},
          doi = {10.1088/0004-637X/759/2/144},
archivePrefix = {arXiv},
       eprint = {1209.5286},
 primaryClass = {astro-ph.SR},
       adsurl = {https://ui.adsabs.harvard.edu/abs/2012ApJ...759..144T}
}

@ARTICLE{2016A&A...Nakariakov,
       author = {{Nakariakov}, V.~M. and {Anfinogentov}, S.~A. and {Nistic{\`o}}, G. and {Lee}, D.-H.},
        title = "{Undamped transverse oscillations of coronal loops as a self-oscillatory process}",
      journal = {\aap},
         year = 2016,
        month = jun,
       volume = {591},
          eid = {L5},
        pages = {L5},
          doi = {10.1051/0004-6361/201628850},
       adsurl = {https://ui.adsabs.harvard.edu/abs/2016A&A...591L...5N}
}

@ARTICLE{2024A&A...Lim,
       author = {{Lim}, Daye and {Van Doorsselaere}, Tom and {Berghmans}, David and {Petrova}, Elena},
        title = "{Characteristics and energy flux distributions of decayless transverse oscillations depending on coronal regions}",
      journal = {\aap},
         year = 2024,
        month = sep,
       volume = {689},
          eid = {A16},
        pages = {A16},
          doi = {10.1051/0004-6361/202450433},
archivePrefix = {arXiv},
       eprint = {2406.06368},
 primaryClass = {astro-ph.SR},
       adsurl = {https://ui.adsabs.harvard.edu/abs/2024A&A...689A..16L}
}

@ARTICLE{2024SoPh..299...WangYM,
       author = {{Wang}, Y.-M.},
        title = "{Coronal Holes, Footpoint Reconnection, and the Origin of the Slow (and Fast) Solar Wind}",
      journal = {\solphys},
         year = 2024,
        month = apr,
       volume = {299},
       number = {4},
          eid = {54},
        pages = {54},
          doi = {10.1007/s11207-024-02300-3},
       adsurl = {https://ui.adsabs.harvard.edu/abs/2024SoPh..299...54W}
}

@ARTICLE{2020ApJ...893...Morgan,
       author = {{Morgan}, Huw and {Cook}, Anthony C.},
        title = "{The Width, Density, and Outflow of Solar Coronal Streamers}",
      journal = {\apj},
         year = 2020,
        month = apr,
       volume = {893},
       number = {1},
          eid = {57},
        pages = {57},
          doi = {10.3847/1538-4357/ab7e32},
archivePrefix = {arXiv},
       eprint = {2003.04809},
 primaryClass = {astro-ph.SR},
       adsurl = {https://ui.adsabs.harvard.edu/abs/2020ApJ...893...57M}
}

@ARTICLE{2024ApJ...973...CS,
       author = {{Kumar}, Pankaj and {Karpen}, Judith T. and {Yurchyshyn}, Vasyl and {DeVore}, C. Richard and {Antiochos}, Spiro K.},
        title = "{Direct Imaging of a Prolonged Plasma/Current Sheet and Quasiperiodic Magnetic Reconnection on the Sun}",
      journal = {\apj},
         year = 2024,
        month = oct,
       volume = {973},
       number = {2},
          eid = {74},
        pages = {74},
          doi = {10.3847/1538-4357/ad63a2},
archivePrefix = {arXiv},
       eprint = {2407.07687},
 primaryClass = {astro-ph.SR},
       adsurl = {https://ui.adsabs.harvard.edu/abs/2024ApJ...973...74K}
}

@ARTICLE{2018ApJ...ChengXin,
       author = {{Cheng}, X. and {Li}, Y. and {Wan}, L.~F. and {Ding}, M.~D. and {Chen}, P.~F. and {Zhang}, J. and {Liu}, J.~J.},
        title = "{Observations of Turbulent Magnetic Reconnection within a Solar Current Sheet}",
      journal = {\apj},
         year = 2018,
        month = oct,
       volume = {866},
       number = {1},
          eid = {64},
        pages = {64},
          doi = {10.3847/1538-4357/aadd16},
archivePrefix = {arXiv},
       eprint = {1808.06071},
 primaryClass = {astro-ph.SR},
       adsurl = {https://ui.adsabs.harvard.edu/abs/2018ApJ...866...64C}
}

@ARTICLE{2019ApJ...French,
       author = {{French}, Ryan J. and {Judge}, Philip G. and {Matthews}, Sarah A. and {van Driel-Gesztelyi}, Lidia},
        title = "{Spectropolarimetric Insight into Plasma Sheet Dynamics of a Solar Flare}",
      journal = {\apjl},
         year = 2019,
        month = dec,
       volume = {887},
       number = {2},
          eid = {L34},
        pages = {L34},
          doi = {10.3847/2041-8213/ab5d34},
archivePrefix = {arXiv},
       eprint = {1911.12666},
 primaryClass = {astro-ph.SR},
       adsurl = {https://ui.adsabs.harvard.edu/abs/2019ApJ...887L..34F}
}

@article{gao2026high,
  title={High-Frequency Magnetohydrodynamic Waves with Substantial Energy in the Solar Polar Corona},
  author={Gao, Yuhang and Tian, Hui and Morton, Richard and Van Doorsselaere, Tom and Lim, Daye and Guo, Mingzhe and He, Jiansen and Hou, Zhenyong},
  journal={National Science Review},
  pages={nwag370},
  year={2026},
  publisher={Oxford University Press}
}

@ARTICLE{2024ApJ...Jarolim,
       author = {{Jarolim}, Robert and {Tremblay}, Benoit and {Mu{\~n}oz-Jaramillo}, Andr{\'e}s and {Bintsi}, Kyriaki-Margarita and {Jungbluth}, Anna and {Santos}, Miraflor and {Vourlidas}, Angelos and {Mason}, James P. and {Sundaresan}, Sairam and {Downs}, Cooper and {Caplan}, Ronald M.},
        title = "{SuNeRF: 3D Reconstruction of the Solar EUV Corona Using Neural Radiance Fields}",
      journal = {\apjl},
         year = 2024,
        month = feb,
       volume = {961},
       number = {2},
          eid = {L31},
        pages = {L31},
          doi = {10.3847/2041-8213/ad12d2},
archivePrefix = {arXiv},
       eprint = {2401.16388},
 primaryClass = {astro-ph.SR},
       adsurl = {https://ui.adsabs.harvard.edu/abs/2024ApJ...961L..31J}
}

@ARTICLE{2024SciA...Schad,
       author = {{Schad}, Thomas A. and {Petrie}, Gordon and {Kuhn}, Jeffrey and {Fehlmann}, Andre and {Rimmele}, Thomas and {Tritschler}, Alexandra and {Woeger}, Friedrich and {Scholl}, Isabelle and {Williams}, Rebecca and {Harrington}, David and {Paraschiv}, Alin and {Szente}, Judit},
        title = "{Mapping the Sun's coronal magnetic field using the Zeeman effect}",
      journal = {Science Advances},
         year = 2024,
        month = sep,
       volume = {10},
       number = {37},
        pages = {eadq1604},
          doi = {10.1126/sciadv.adq1604},
archivePrefix = {arXiv},
       eprint = {2410.21568},
 primaryClass = {astro-ph.SR},
       adsurl = {https://ui.adsabs.harvard.edu/abs/2024SciA...10.1604S}
}

@ARTICLE{2026ApJ..Wraback,
       author = {{Wraback}, Elizabeth M. and {Farid}, Samaiyah and {Schad}, Thomas A. and {B{\k{a}}k-St{\c{e}}{\'s}licka}, Urszula and {de Toma}, Giuliana and {Fan}, Yuhong and {Gibson}, Sarah E. and {Paz Miralles}, Mari and {Zhang}, Jie and {Dhakal}, Suman K.},
        title = "{Coronal Cavity Magnetic Fields Inferred by DKIST/CryoNIRSP}",
      journal = {\apjl},
         year = 2026,
        month = jul,
       volume = {1005},
       number = {1},
          eid = {L8},
        pages = {L8},
          doi = {10.3847/2041-8213/ae7b33},
       adsurl = {https://ui.adsabs.harvard.edu/abs/2026ApJ..1005L...8W}
}

@ARTICLE{2016JGRA..UCoMP,
       author = {{Landi}, E. and {Habbal}, S.~R. and {Tomczyk}, S.},
        title = "{Coronal plasma diagnostics from ground-based observations}",
      journal = {Journal of Geophysical Research (Space Physics)},
         year = 2016,
        month = sep,
       volume = {121},
       number = {9},
        pages = {8237-8249},
          doi = {10.1002/2016JA022598},
       adsurl = {https://ui.adsabs.harvard.edu/abs/2016JGRA..121.8237L}
}

@ARTICLE{2007Sci...317.1192T,
       author = {{Tomczyk}, S. and {McIntosh}, S.~W. and {Keil}, S.~L. and {Judge}, P.~G. and {Schad}, T. and {Seeley}, D.~H. and {Edmondson}, J.},
        title = "{Alfv{\'e}n Waves in the Solar Corona}",
      journal = {Science},
         year = 2007,
        month = aug,
       volume = {317},
       number = {5842},
        pages = {1192-1196},
          doi = {10.1126/science.1143304},
       adsurl = {https://ui.adsabs.harvard.edu/abs/2007Sci...317.1192T}
}

@ARTICLE{2022ApJ...930...Gao,
       author = {{Gao}, Yuhang and {Tian}, Hui and {Van Doorsselaere}, Tom and {Chen}, Yajie},
        title = "{Decayless Oscillations in Solar Coronal Bright Points}",
      journal = {\apj},
         year = 2022,
        month = may,
       volume = {930},
       number = {1},
          eid = {55},
        pages = {55},
          doi = {10.3847/1538-4357/ac62cf},
archivePrefix = {arXiv},
       eprint = {2203.17034},
 primaryClass = {astro-ph.SR},
       adsurl = {https://ui.adsabs.harvard.edu/abs/2022ApJ...930...55G}
}

@ARTICLE{2001A&A...372L..53N,
       author = {{Nakariakov}, V.~M. and {Ofman}, L.},
        title = "{Determination of the coronal magnetic field by coronal loop oscillations}",
      journal = {\aap},
         year = 2001,
        month = jun,
       volume = {372},
        pages = {L53-L56},
          doi = {10.1051/0004-6361:20010607},
       adsurl = {https://ui.adsabs.harvard.edu/abs/2001A&A...372L..53N}
}

@ARTICLE{2023SoPh..298..Liu,
       author = {{Liu}, X. and {Tian}, H. and {T{\"o}r{\"o}k}, T. and {Gibson}, S. and {Yang}, Z. and {Li}, W. and {Samanta}, T.},
        title = "{Magnetic Field and Plasma Diagnostics for Solar Coronal Mass Ejections: A Case Study Using the Forward Modeling Approach}",
      journal = {\solphys},
         year = 2023,
        month = oct,
       volume = {298},
       number = {10},
          eid = {112},
        pages = {112},
          doi = {10.1007/s11207-023-02207-5},
       adsurl = {https://ui.adsabs.harvard.edu/abs/2023SoPh..298..112L}
}

@ARTICLE{2009A&A...495..Young,
       author = {{Young}, P.~R. and {Watanabe}, T. and {Hara}, H. and {Mariska}, J.~T.},
        title = "{High-precision density measurements in the solar corona. I. Analysis methods and results for Fe XII and Fe XIII}",
      journal = {\aap},
         year = 2009,
        month = feb,
       volume = {495},
       number = {2},
        pages = {587-606},
          doi = {10.1051/0004-6361:200810143},
archivePrefix = {arXiv},
       eprint = {0805.0958},
 primaryClass = {astro-ph},
       adsurl = {https://ui.adsabs.harvard.edu/abs/2009A&A...495..587Y}
}

@ARTICLE{2013PASJ...65S..14I,
       author = {{Iwai}, Kazumasa and {Shibasaki}, Kiyoto},
        title = "{Measurements of Coronal and Chromospheric Magnetic Fields using Polarization Observations by the Nobeyama Radioheliograph}",
      journal = {\pasj},
         year = 2013,
        month = dec,
       volume = {65},
          eid = {S14},
        pages = {S14},
          doi = {10.1093/pasj/65.sp1.S14},
archivePrefix = {arXiv},
       eprint = {1308.2283},
 primaryClass = {astro-ph.SR},
       adsurl = {https://ui.adsabs.harvard.edu/abs/2013PASJ...65S..14I}
}

@ARTICLE{2018ApJ...856..Magyar,
       author = {{Magyar}, N. and {Van Doorsselaere}, T.},
        title = "{Assessing the Capabilities of Dynamic Coronal Seismology of Alfv{\'e}nic Waves through Forward Modeling}",
      journal = {\apj},
         year = 2018,
        month = apr,
       volume = {856},
       number = {2},
          eid = {144},
        pages = {144},
          doi = {10.3847/1538-4357/aab42c},
archivePrefix = {arXiv},
       eprint = {1804.02175},
 primaryClass = {astro-ph.SR},
       adsurl = {https://ui.adsabs.harvard.edu/abs/2018ApJ...856..144M}
}

@ARTICLE{2026NatAs..10...42M,
       author = {{Morton}, R.~J. and {Gao}, Y. and {Tajfirouze}, E. and {Tian}, H. and {Van Doorsselaere}, T. and {Schad}, T.~A.},
        title = "{Evidence for small-scale torsional Alfv{\'e}n waves in the solar corona}",
      journal = {Nature Astronomy},
         year = 2026,
        month = jan,
       volume = {10},
        pages = {42-53},
          doi = {10.1038/s41550-025-02690-9},
       adsurl = {https://ui.adsabs.harvard.edu/abs/2026NatAs..10...42M}
}

@ARTICLE{2025ApJS..280...48L,
       author = {{Luo}, Hongyang and {Fan}, Yuhong},
        title = "{The Yin─Yang Magnetic Flux Eruption (Yin─Yang MFE) Code: A Global Corona Magnetohydrodynamic Code with the Yin─Yang Grid}",
      journal = {\apjs},
         year = 2025,
        month = oct,
       volume = {280},
       number = {2},
          eid = {48},
        pages = {48},
          doi = {10.3847/1538-4365/adf8d1},
archivePrefix = {arXiv},
       eprint = {2508.08210},
 primaryClass = {astro-ph.SR},
       adsurl = {https://ui.adsabs.harvard.edu/abs/2025ApJS..280...48L}
}

@ARTICLE{2022ApJ...941...61F,
       author = {{Fan}, Yuhong},
        title = "{An Improved Magnetohydrodynamic Simulation of the 2006 December 13 Coronal Mass Ejection of NOAA Active Region 10930}",
      journal = {\apj},
         year = 2022,
        month = dec,
       volume = {941},
       number = {1},
          eid = {61},
        pages = {61},
          doi = {10.3847/1538-4357/aca0ec},
archivePrefix = {arXiv},
       eprint = {2211.03736},
 primaryClass = {astro-ph.SR},
       adsurl = {https://ui.adsabs.harvard.edu/abs/2022ApJ...941...61F}
}

@ARTICLE{2017ApJ...844...26F,
       author = {{Fan}, Yuhong},
        title = "{MHD Simulations of the Eruption of Coronal Flux Ropes under Coronal Streamers}",
      journal = {\apj},
         year = 2017,
        month = jul,
       volume = {844},
       number = {1},
          eid = {26},
        pages = {26},
          doi = {10.3847/1538-4357/aa7a56},
archivePrefix = {arXiv},
       eprint = {1706.06076},
 primaryClass = {astro-ph.SR},
       adsurl = {https://ui.adsabs.harvard.edu/abs/2017ApJ...844...26F}
}

@ARTICLE{2016JGRA..121.7470T,
       author = {{Tomczyk}, S. and {Landi}, E. and {Burkepile}, J.~T. and {Casini}, R. and {DeLuca}, E.~E. and {Fan}, Y. and {Gibson}, S.~E. and {Lin}, H. and {McIntosh}, S.~W. and {Solomon}, S.~C. and {de Toma}, G. and {de Wijn}, A.~G. and {Zhang}, J.},
        title = "{Scientific objectives and capabilities of the Coronal Solar Magnetism Observatory}",
      journal = {Journal of Geophysical Research (Space Physics)},
         year = 2016,
        month = aug,
       volume = {121},
       number = {8},
        pages = {7470-7487},
          doi = {10.1002/2016JA022871},
       adsurl = {https://ui.adsabs.harvard.edu/abs/2016JGRA..121.7470T}
}

@ARTICLE{2026A&A...710A.138A,
       author = {{Amato}, E. and {Bemporad}, A. and {Susino}, R. and {Frassati}, F. and {De Leo}, Y. and {Bergamin}, G. and {Andretta}, V. and {Fineschi}, S. and {Jerse}, G. and {Landini}, F. and {Nicolini}, G. and {Pancrazzi}, M. and {Russano}, G. and {Sasso}, C. and {Teriaca}, L. and {Uslenghi}, M.},
        title = "{Investigating the formation of post-coronal mass ejection plasmoids observed by the Metis coronagraph on board Solar Orbiter}",
      journal = {\aap},
         year = 2026,
        month = jun,
       volume = {710},
          eid = {A138},
        pages = {A138},
          doi = {10.1051/0004-6361/202555747},
       adsurl = {https://ui.adsabs.harvard.edu/abs/2026A&A...710A.138A}
}
\bibliographystyle{aasjournalv7}

\end{CJK*}
\end{document}